\documentclass[10pt]{iopart}
\usepackage{iopams} 
\usepackage{graphicx}

\usepackage{mcite}
\usepackage{cite}
\usepackage{multirow}
\usepackage[caption=false]{subfig}
\begin{document}

\title{Self-Learning Kinetic Monte Carlo Simulations of Al Diffusion in Mg}

\author{Giridhar Nandipati$^1$, Niranjan Govind, Amity Andersen, Aashish Rohatgi$^2$}
\ead{$^1$giridhar.nandipati@pnnl.gov}
\ead{$^2$aashish.rohatgi@pnnl.gov} 	
\address{Pacific Northwest National Laboratory, Richland, WA, 99352, USA}

\date{\today}

\begin{abstract}
Vacancy-mediated diffusion of an Al atom in pure Mg matrix is studied using the atomistic, on-lattice self-learning kinetic Monte Carlo (SLKMC) method.  Activation barriers for vacancy-Mg and vacancy-Al atom exchange processes are calculated on-the-fly using the climbing image nudged-elastic band method and binary Mg-Al modified embedded-atom method interatomic potential.  Diffusivities of an Al atom obtained from SLKMC simulations show the same behavior as observed in experimental and theoretical studies available in the literature, that is, Al atom diffuses faster within the basal plane than along the $c-$axis. Although, the effective activation barriers for Al-atom diffusion from SLKMC simulations are close to experimental and theoretical values,  the effective prefactors are lower than those obtained from experiments.
We present all the possible vacancy-Mg and vacancy-Al atom exchange processes and their activation barriers identified in SLKMC simulations. 
A simple mapping scheme to map an HCP lattice on to a simple cubic lattice is described, which enables the simulation of HCP lattice using on-lattice framework. We also present the pattern recognition scheme which is used in SLKMC simulations to identify the local Al atom configuration around a vacancy.
\end{abstract}
\maketitle


\section{Introduction}	

This study is motivated by an effort to understand the microstructural evolution and kinetics in Mg-Al based alloys during solidification, especially during the subsequent heat-treatment.  Mg alloys are attractive for engineering applications on account of their light weight.\cite{Mg_ref_1, Mg_ref_2,Mg_ref_3} 
Automotive industry is increasingly considering Mg alloys for light-weighting cars and trucks to achieve better fuel efficiency and hence, meet the government mandated fuel economy and emission standards.
Mg-Al based alloys are also attractive because aluminum increases the strength and widens the solidification range, thus, making Mg-Al alloys easy to cast into shapes required for automotive applications.  However, kinetics of underlying atomistic processes  of microstructural evolution in Mg alloys is not fully known or understood.  Moreover the knowledge of reliable Al diffusivities in Mg-Al alloys is necessary for higher-scale models that can reach engineering length and time scales.
Therefore, it is our expectation that an improved predictive capability, enabled by the atomistic modeling work presented here will be valuable in improving the design, manufacture and processing of Mg alloys.

Modeling of diffusion related phenomena in alloys is particularly challenging because it largely consists of an interplay of ensemble of thermally activated atomic jumps whose activation barriers depend on local atomic arrangement. 
In general, the challenge is to relate diffusion constants to atomic-scale processes and energetics.
Depending on the temperature and complexity, it is possible to study the kinetics of phase separation spanning time scales of minutes to hours using the kinetic Monte Carlo (KMC) method. \cite{kristen,kmc1} 
Kinetics of coherent phase transformation have been successfully studied primarily using various broken bond models in fcc and bcc alloys using KMC method.\cite{kmc_seg_1, kmc_seg_2, kmc_seg_3,kmc_seg_4,kmc_seg_5,kmc_seg_6,kmc_seg_7,kmc_seg_8,kmc_seg_9,kmc_seg_10,kmc_seg_11,kmc_seg_12}
To the best of our knowledge, no such simulations were performed in HCP alloys.
This paper focuses on the development and our implementation of the SLKMC method\cite{slkmc1} and its application to the study of vacancy and Al atom diffusion in Mg matrix.
  
This paper is organized as follows. In Section~\ref{slkmc}, we describe very briefly the SLKMC method. In Section.~\ref{mapping} and~\ref{pattern} we describe in detail the mapping scheme used to map the HCP lattice onto the simple cubic lattice and the pattern recognition scheme to identify local Al-atom neighborhood around a vacancy.  In Section~\ref{results}, we present simulation results for a vacancy diffusion in pure Mg and  an Al atom diffusion in pure single crystal Mg. We then present a comparison of Al atom diffusivities obtained from SLKMC simulations with experimental and theoretical values available in the literature. In Section~\ref{AlMgx}, we discuss some of the vacancy-Mg and vacancy-Al atom exchange processes and their activations barriers. In addition, we present all vacancy-Mg and vacancy-Al atom exchange processes and their activation barriers for Al atom diffusion, identified in SLKMC simulations. Finally, in Section.~\ref{conclusions}, we summarize our results.
Note that the details on the parameters of the MEAM interatomic potential are provided in the supplemental information.

\section{Self-learning Kinetic Monte Carlo Method}\label{slkmc}

The SLKMC method \cite{slkmc1} is a variant of on-the-fly KMC method,\cite{slkmc2}  which identifies all the possible processes and calculates their activation energies based on the local atomic neighborhood on-the-fly. It uses a pattern-recognition scheme to store and retrieve those identified processes and their activation energies from a database to avoid redundant calculations.
A pattern-recognition scheme \cite{slkmc1,a_slkmc2,a_slkmc5, a_slkmc7, a_slkmc8} is used to generate a unique tag or an identifier for a particular arrangement of neighboring atoms (or in short, `configuration') around an atom. All the identified diffusion processes the atom can perform along with their activation barriers are attached to the unique tag and are stored in a database. Note that the KMC algorithm we used is also based on the so-called ``n-fold-way'' or Bortz-Kalos-Lebowitz(BKL)\cite{bkl} or the \textit{residence-time} algorithm.

\section{Mapping of HCP lattice on to a cubic lattice}\label{mapping}

\begin{figure*}
  \centering
    \subfloat[(1) two A- and B- layers (2) transformed lattice and where each sphere represents two atoms]{
  \includegraphics[width=2.45in]{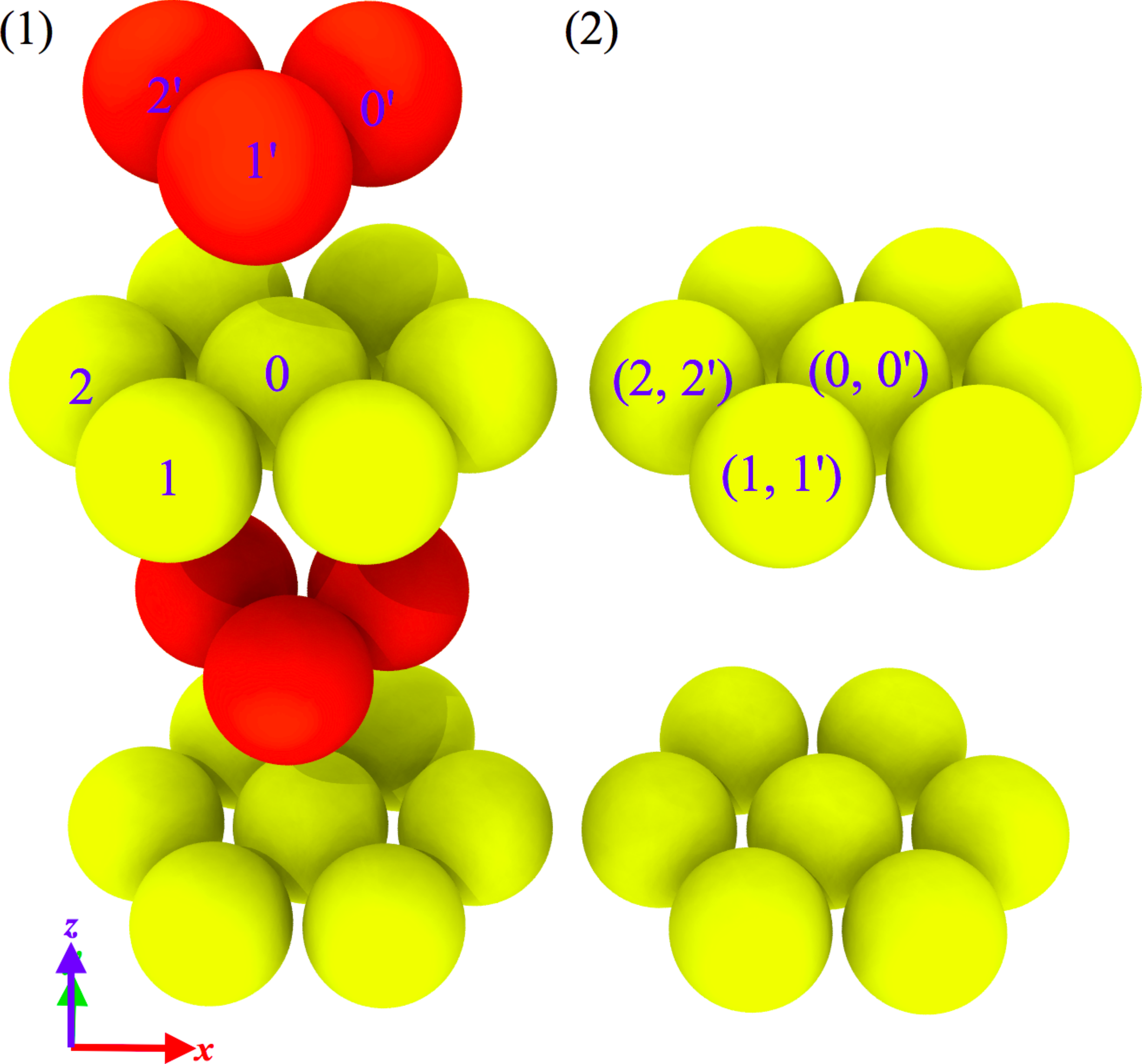}
  \label{fig1}
  }
    \hspace{5pt}
      \subfloat[Mapping of cubic lattice coordinates on to a HCP lattice]{
  \includegraphics[width=4.00in]{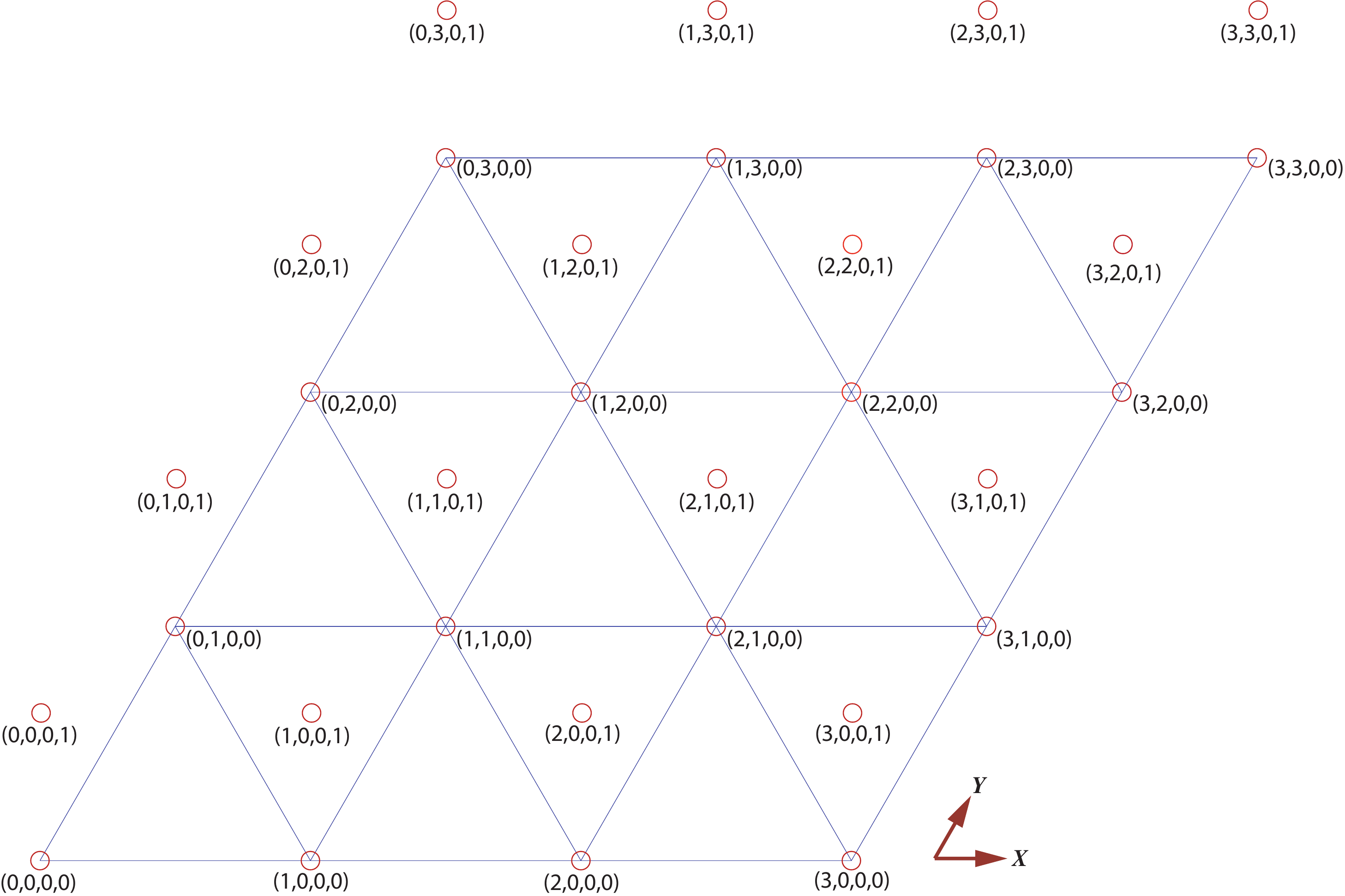}
  \label{fig2}
  }
  \caption{(a) Transformation of the HCP lattice from a non-Bravais into a Bravais lattice (b) Mapping of cubic lattice coordinates on to the hexagonal lattice}
\end{figure*}

To use an integer coordinate system in on-lattice KMC simulations, the crystal structure of a material should be mapped onto the simple cubic lattice.
Since HCP lattice is a non-Bravais lattice, it has to be first converted into a Bravais lattice before the mapping. Note that an HCP lattice consists of two interpenetrating 2D hexagonal layers called A and B layers as shown in Fig. (\ref{fig1}-1).  By combining A \& B layers into a single layer, the HCP lattice is transformed into a stack of 2D hexagonal planes with each lattice point representing two atoms (one from the A-layer and another from the B-layer) as shown in Fig. (\ref{fig1}-2) and with an inter-layer spacing equal to $c$ (= 5.21 \AA). 
A cubic lattice with a basis requires four integer coordinates, $x, ~y, ~z$ and $b$ (basis). For an HCP lattice, $b$ = 0 or 1 indicates whether the atom is in the A or B layer, respectively. Fig.~\ref{fig2} shows an illustration of the mapping of integer coordinate system on to a hexagonal lattice. Coordinates of each atom in the real space are obtained from the integer coordinates using the following expressions:
\begin{eqnarray}
x_{real} &=  a_0 \bigg(x_{int} + \frac{y_{int}}{2}\bigg)\nonumber\\[0.5em]
y_{real} &= \frac{a_0}{\sqrt(3)}\bigg(\frac{3}{2}~y_{int} + b\bigg) \nonumber \\[0.5em]
z_{real} &= c\bigg(z_{int}+\frac{b}{2}\bigg)\nonumber
\end{eqnarray}
where $x_{int}, ~y_{int}, ~z_{int}, ~b$ are integer coordinates in the cubic lattice and $x_{real}, ~y_{real}, ~z_{real}$ are the coordinates in real space. $a_0$ and $c$ are HCP lattice constants.

\section{Pattern Recognition Scheme}\label{pattern}

\begin{figure}
	\centering
	\includegraphics[width=1.0in]{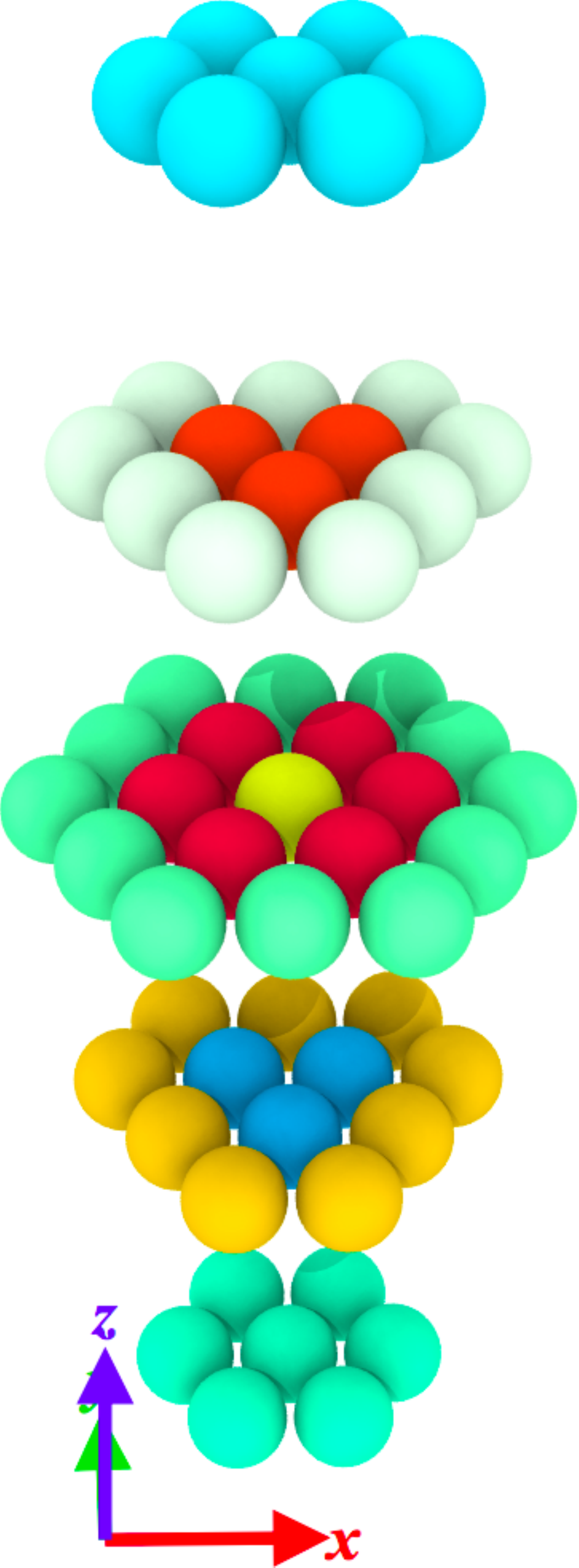}
	\caption{56 neighborhood atoms around the central vacancy (greenish yellow sphere in the center of middle layer). Atoms that belong to a ring are given the same color. }
	\label{fig3}
	
\end{figure}

\begin{figure*}
	\centering
	\includegraphics[width=6.7in]{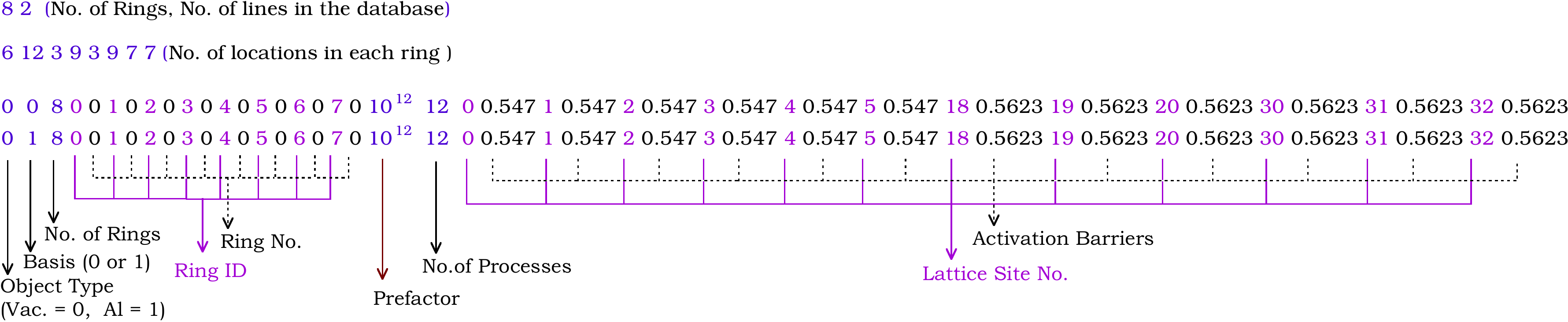}
	\caption{Format of the database. }
	\label{dbformat}
	
\end{figure*}

To properly include their effect, Al atoms in the 56 neighborhood sites around a vacancy (greenish-yellow sphere at the center of the middle layer) as shown in Fig.~\ref{fig3} are included in the calculation of activation barriers for vacancy-Mg and vacancy-Al atom exchange processes. A unique binary bit pattern is generated for each local Al-atom configuration (i.e., the arrangement of Al atoms around a vacancy) based on the occupancy of only Al atoms. That is, the presence of an Al atom at a site is taken as 1 and, 0 otherwise. For simplicity's sake, 56 sites are grouped into 8 rings as shown in Fig.~\ref{fig3}. Accordingly, each configuration is given by a set of 8  decimal numbers corresponding to the bit patterns of the 8 rings, from hereon are referred to as ring numbers. For each Al-atom configuration, the set of ring numbers along with vacancy-Mg and vacancy-Al atom exchange processes and their activation barriers are stored in a database in the format shown in Fig.~\ref{dbformat}.

\begin{figure}
	\centering
	\includegraphics[width=1.80in]{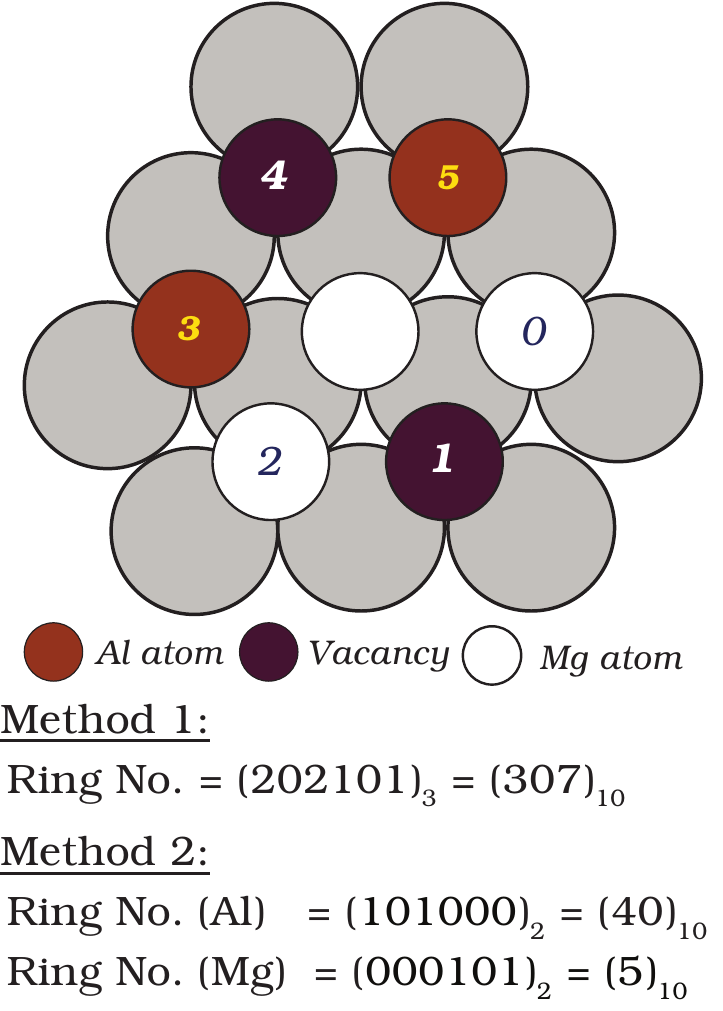}
	\caption{Alternate pattern recognition schemes to identify the neighborhood of an Mg atom. (atoms on the top layer are shown as smaller circles)}
	\label{figp}
	
\end{figure}

Note that if the local neighborhood around an Mg atom is used instead, then each neighborhood lattice site has 3 states; 0, 1 and 2, representing a vacancy, an Mg atom and an Al atom, respectively. As shown in Fig.~\ref{figp}, one can either use a single decimal ring number for each ring corresponding to the ternary (radix-3 or base-3) number \cite{radix3} generated based on Mg and Al atom occupancy, or two decimal numbers for each ring corresponding to the two binary bit patterns based on Al and Mg atom occupancy. To make the database concise we used the set of ring numbers which represent the local Al-atom neighborhood around a vacancy (a non-real object) as described in the previous paragraph. 

In present simulations a vacancy in the HCP lattice is only allowed to hop to 12 nearest neighbors (NN) sites; 6 NN are within the basal plane and 6 NN are in the adjacent basal planes (3 each in the plane above and below). Local Al-atom configuration dependent activation barriers for a vacancy hop (jump) or vacancy-atom exchange processes are calculated using the climbing image nudge-elastic band (CI-NEB) method \cite{CINEB} as implemented in LAMMPS, an MD simulation package. \cite{LAMMPS} For the CI-NEB calculations, we used a periodic orthorhombic simulation cell with 512 hcp lattice points and consists a sum of 511 Mg and Al atoms, and one vacancy. Al atoms are placed at lattice sites around the vacancy according to the Al-atom configuration not found in the database and rest of the sites are occupied by Mg atoms. Note that the number of Mg and Al atoms varies based on the unknown Al-configuration, but their sum is always equal to 511 atoms.

 During the course of a simulation if an Al-atom configuration  or its symmetric equivalent is not found in the database, then the CI-NEB module within the LAMMPS is invoked. To reduce the number of CI-NEB calculations, we exploited the six-fold symmetry of the HCP lattice along the basal plane and mirror symmetry along the $c-$axis. The following eight unique symmetry operations were used to recognize equivalent configurations: (1) $120^{\circ}$ ($\theta_{\mathrm{120}}^c$) and (2) $240^{\circ}$ ($\theta_{\mathrm{240}}^c$) rotation around the $c-$axis, mirror reflection in (3) $xy-$($R_{xy}$) and (4) $yz-$($R_{xy}$) planes (Fig.~\ref{fig1}) (5) $\theta_{\mathrm{120}}^c$ and $R_{xy}$ (6) $\theta_{\mathrm{120}}^c$ and $R_{yz}$ (7) $\theta_{\mathrm{240}}^c$ and $R_{xy}$ (8) $\theta_{\mathrm{240}}^c$ and $R_{yz}$. Note that the choice of using Al-atom environment around a vacancy also reduces the frequency of usage of symmetry operations during a simulation.

For the inter-atomic interactions, we used a second nearest neighbor (2NN) modified embedded-atom method (MEAM) potential as developed by Lee {\it et al}. \mcite{*meam1, *meam2}
The original MEAM potential parameters for Mg-Al systems were taken from Ref.~\cite{mg_meam1} and were adjusted to get material parameters of pure Mg and Al, Al in Mg and Mg in Al, and Mg$_{17}$Al$_{12}$ close to first-principles and experimental values available in the literature. Modified MEAM parameters and comparison of material parameters with those obtained using density functional theory (DFT) calculations and from experiments are given in the supplemental information.

The mapping method described earlier is implemented in a general way such that different types of lattice structures can be mapped onto a simple cubic lattice, as part of a larger KMC simulation package, AKSOME (Atomistic Kinetic Simulations of Microstructural evolution). All the required input, which includes locations of neighborhood sites in each ring relative to vacancy location, symmetry operations and simulation parameters are provided via input files.  AKSOME is a flexible, on-lattice, self-learning kinetic Monte Carlo code being developed at PNNL, designed to study defect diffusion and segregation in multi-component alloys.
\footnote{AKSOME is still under development, therefore some aspects (e.g. object type, number of rings being specified twice, ring ID) of the format of the database shown in Fig.~\ref{dbformat} are redundant for present simulations, but are needed for the additional functionality not described in the present article. }

\begin{table*}
	\footnotesize
	\caption{Comparison of vacancy diffusivities within a basal plane and along the $c-$axis  in pure Mg obtained from KMC simulations and those calculated using Eqs.~\ref{p0}-\ref{dtotal} at 300, 400 and 500 K.}
	\begin{center}
		\begin{tabular}{@{}ccccccc} 
			\br
			\multirow{3}{*}{Temperature (K)} & \multicolumn{2}{c}{$D_{\parallel}$ (\AA$^2/s$)}&\multicolumn{2}{c}{$D_{\perp}$(\AA$^2/s$)} & \multicolumn{2}{c}{$D_{total}$(\AA$^2/s$)}\\[1ex]
			\cline{2-7}
			& \multicolumn{1}{c}{}&\multicolumn{1}{c|}{}&\multicolumn{1}{c}{} & \multicolumn{1}{c|}{} & \multicolumn{1}{c}{}\\
			\ns
			& \multicolumn{1}{c}{KMC}&\multicolumn{1}{c|}{Eq.~\ref{p0}}&\multicolumn{1}{c}{KMC} & \multicolumn{1}{c|}{Eq.~\ref{p1}} & \multicolumn{1}{c}{KMC} & \multicolumn{1}{c}{Eq.~\ref{dtotal}}\\[0.5ex]
			\hline 
			& \multicolumn{1}{c}{}&\multicolumn{1}{c|}{}&\multicolumn{1}{c}{} & \multicolumn{1}{c|}{} & \multicolumn{1}{c}{}\\
			\ns
			300 & \multicolumn{1}{c}{1.43$\times 10^5$} & \multicolumn{1}{c|}{1.45$\times 10^5$} & \multicolumn{1}{c}{2.32$\times 10^5$ } & \multicolumn{1}{c|}{2.36$\times 10^5$} & \multicolumn{1}{c}{2.06$\times 10^5$ } & \multicolumn{1}{c}{2.12$\times 10^5$}\\[0.5ex]
			400 & \multicolumn{1}{c}{3.40$\times 10^7$} & \multicolumn{1}{c|}{3.32$\times 10^7$} & \multicolumn{1}{c}{4.80$\times 10^7$ } & \multicolumn{1}{c|}{4.80$\times 10^7$}&\multicolumn{1}{c}{4.30$\times 10^7$ } & \multicolumn{1}{c}{4.32$\times 10^7$}\\[0.5ex]
			500 & \multicolumn{1}{c}{8.80$\times 10^8$} & \multicolumn{1}{c|}{8.70$\times 10^8$} & \multicolumn{1}{c}{1.18$\times 10^9$ } & \multicolumn{1}{c|}{1.17$\times 10^9$}&\multicolumn{1}{c}{1.17$\times 10^9$ } & \multicolumn{1}{c}{1.16$\times 10^9$}\\
			\br
		\end{tabular}
		\label{dc}
	\end{center}
\end{table*}

\section{Results and Discussion}\label{results}

We will first describe  simulation details and then determine the diffusivity of a single vacancy in pure Mg. Finally we will determine the diffusivity of an Al atom in the Mg matrix.

\subsection{Simulation Details}
We used a periodic simulation cell of $128 \times 128 \times 128 $ lattice points, which corresponds to $410.88 \times 355.83 \times 664.80$ \AA$^3$ in real space. Note that with a basis of 2 at each lattice point, the simulation cell consists of  $2^{22}$ atoms ($N_a$).  Mg lattice constants in our calculations are taken as $a = 3.21$\AA  ~and $c/a = 1.623$. For this simulation cell size, presence of one vacancy corresponds approximately to the equilibrium vacancy concentration at a temperature of 677.3 K (404 $^{\circ}$C) in pure Mg, which is $1.03 \times 10^{22}$ vacancies/m$^{-3}$. Therefore,  the actual physical time ($t_{real}$), at other temperatures is obtained by rescaling the KMC simulation time ($t_{kmc}$) as \cite{kmc_seg_1, kmc_seg_2}
\begin{eqnarray}
t_{real}= \frac{C_V^{kmc}}{C_V^{eq}}t_{kmc} \label{treal}
\end{eqnarray}
where $C_V^{eq} = \mathrm{exp}(\Delta S^V_f/k_B) \mathrm{exp}(-E^V_f/k_BT)$  is the  equilibrium vacancy concentration at temperature $T$ and $C_V^{kmc} = 1/N_a$ is the vacancy concentration in the simulation cell. $k_B$ is the Boltzmann's constant, $E^{V}_f$ and $\Delta S^V_f$ are the formation energy and entropy of a vacancy, respectively. $E^V_f$ = 0.89 eV, which is calculated using the MEAM potential and $\Delta S_V^f$ = 1.49$k_B$, is taken from the DFT study by Ganeshan {\it et al}.\cite{Mg_diff_dft} 

\begin{figure}
	\centering
	\includegraphics[width=2.45in]{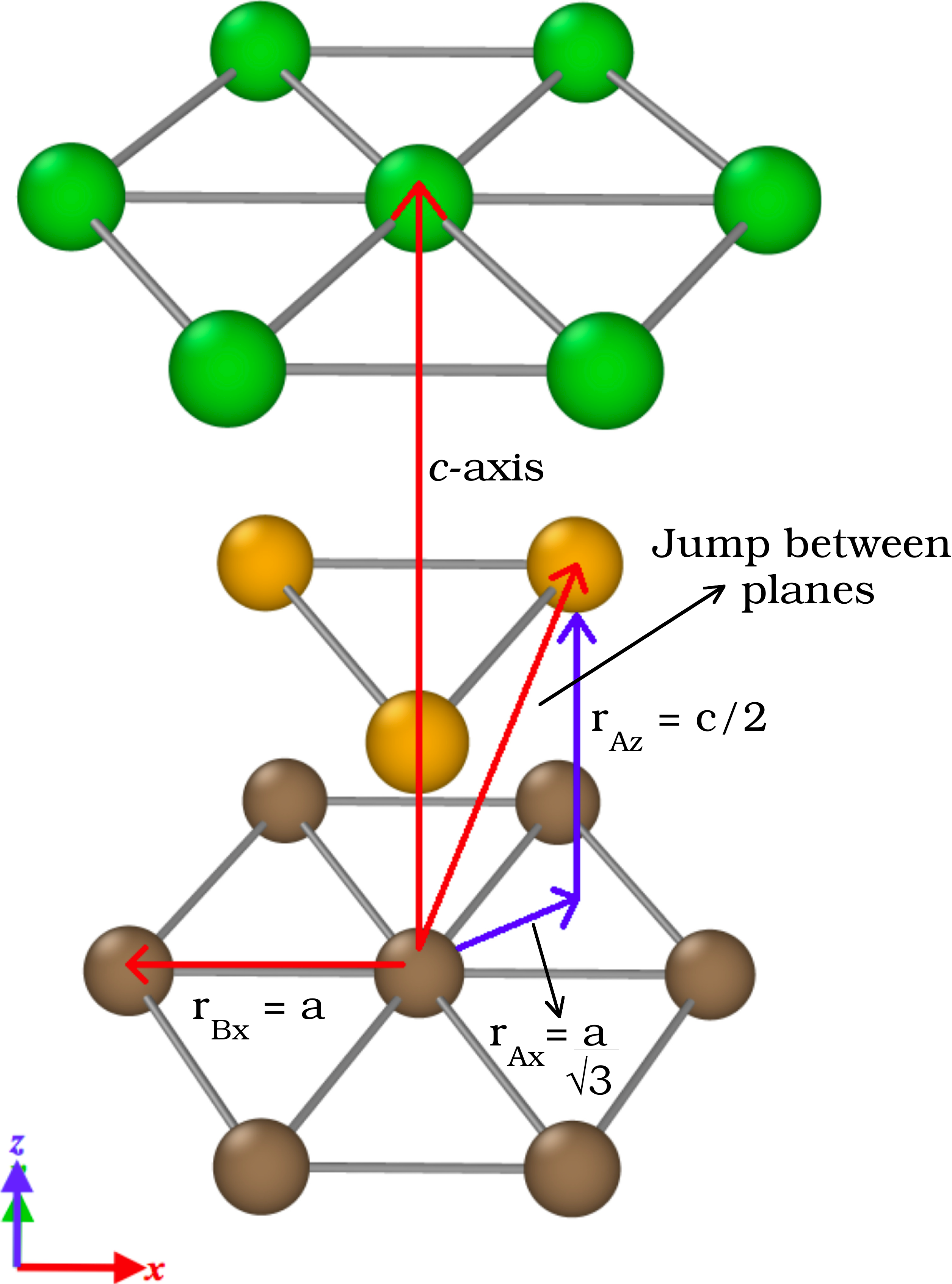}
	\caption{HCP lattice showing jumps and jump distances between basal planes ($r_{Ax}, r_{Az}$) and within basal plane ($r_{Bx}$).}
	\label{hcpfig}
\end{figure}

\subsection{Single Vacancy Diffusion in Pure Mg}

\begin{figure}
	\centering
	\includegraphics[width=2.45in]{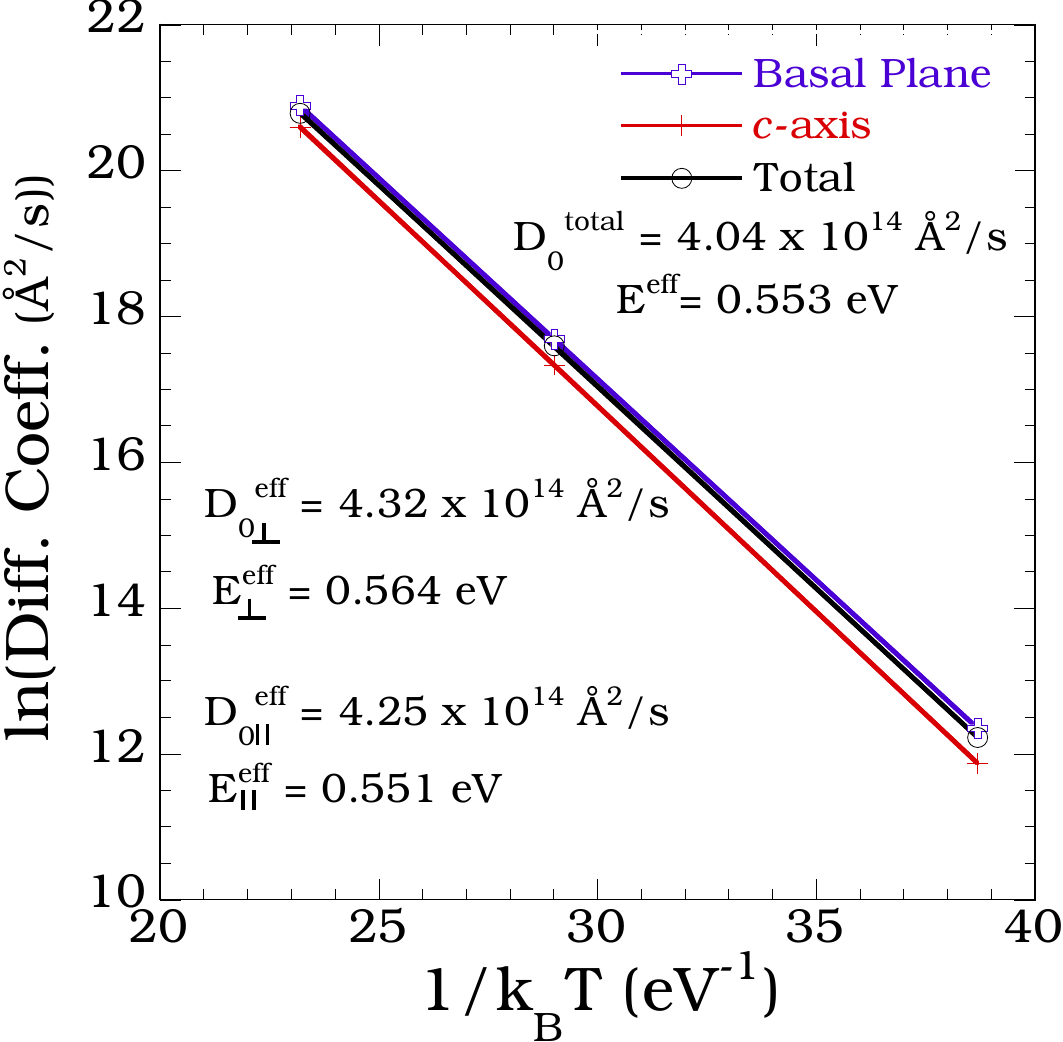}
	\caption{Arrhenius plot of vacancy diffusivities in Mg. Effective energy barrier and prefactors for a vacancy diffusion along ($\parallel$) and perpendicular ($\perp$) to $c-$axis, and total diffusion.}
	\label{fig5}
\end{figure}

\begin{table*}
	\caption{Effective activation energies and diffusion prefactors for an Al atom diffusion in  Mg matrix obtained from SLKMC simulations and a comparison with those obtained from 8-frequency model and experiments}
	\footnotesize
	\begin{center}
%		\footnotesize
\normalsize
		\resizebox{\textwidth}{!}{%
		\addtolength{\tabcolsep}{-4pt}
			\begin{tabular}{@{}lcccccc||cccccc} 
				\br
 & \multicolumn{6}{c||}{Activation Energy (eV)}&\multicolumn{6}{c}{Prefactor (D$_0$)(\AA$^2/s$)}\\[1.5ex]
				\cline{2-13}\cline{2-13}
				& \multicolumn{1}{c}{}&\multicolumn{1}{c}{}&\multicolumn{1}{c|}{} & \multicolumn{1}{c}{} &\multicolumn{1}{c}{}& \multicolumn{1}{c||}{}	& \multicolumn{1}{c}{}&\multicolumn{1}{c}{}&\multicolumn{1}{c|}{} & \multicolumn{1}{c}{} &\multicolumn{1}{c}{}& \multicolumn{1}{c}{}\\ 
				\ns
& & \multicolumn{2}{c|}{Calculations}&\multicolumn{3}{c||}{Experimental}&&\multicolumn{2}{c|}{Calculations}&\multicolumn{3}{c}{Experimental}\\[1ex]
 \cline{3-4} \cline{5-7}\cline{9-13}
 & \multicolumn{1}{c}{}&\multicolumn{1}{c}{}&\multicolumn{1}{c|}{} & \multicolumn{1}{c}{} &\multicolumn{1}{c}{}& \multicolumn{1}{c||}{} & \multicolumn{1}{c}{}&\multicolumn{1}{c}{}&\multicolumn{1}{c|}{} & \multicolumn{1}{c}{} &\multicolumn{1}{c}{}& \multicolumn{1}{c}{}\\ 
				\ns
				& \multicolumn{1}{c}{SLKMC}&\multicolumn{1}{c}{Ganeshan}&\multicolumn{1}{c|}{Zhou}&\multicolumn{1}{c}{Das} & \multicolumn{1}{c}{Kammerer}&\multicolumn{1}{c||}{Brennan} &\multicolumn{1}{c}{SLKMC}&\multicolumn{1}{c}{Ganeshan}&\multicolumn{1}{c|}{Zhou}&\multicolumn{1}{c}{Das} & \multicolumn{1}{c}{Kammerer}&\multicolumn{1}{c}{Brennan}\\[0.5ex]
				& \multicolumn{1}{c}{}&\multicolumn{1}{c}{\cite{Al_diff_dft}}&\multicolumn{1}{c|}{\cite{al_dft_new}}&\multicolumn{1}{c}{\mcite{*all_diff_exp1,*all_diff_exp11}} & \multicolumn{1}{c}{\cite{all_diff_exp2}}&\multicolumn{1}{c||}{\mcite{brennan, Brennen2}} & \multicolumn{1}{c}{} & \multicolumn{1}{c}{\cite{Al_diff_dft}} & \multicolumn{1}{c|}{\cite{al_dft_new}}& \multicolumn{1}{c}{\mcite{*all_diff_exp1,*all_diff_exp11}} & \multicolumn{1}{c}{\cite{all_diff_exp2}}& \multicolumn{1}{c}{\cite{brennan, Brennen2}} \\[0.5ex]
				\hline \hline
				& \multicolumn{1}{c}{}&\multicolumn{1}{c}{}&\multicolumn{1}{c|}{} & \multicolumn{1}{c}{} &\multicolumn{1}{c}{}& \multicolumn{1}{c||}{}& \multicolumn{1}{c}{}&\multicolumn{1}{c}{}&\multicolumn{1}{c|}{} & \multicolumn{1}{c}{} &\multicolumn{1}{c}{}& \multicolumn{1}{c}{}\\
				\ns
				Basal Plane& \multicolumn{1}{c}{1.464} & \multicolumn{1}{c}{1.42}&\multicolumn{1}{c|}{1.34} & \multicolumn{1}{c}{1.60$\pm$0.04}&\multicolumn{1}{c}{-} & \multicolumn{1}{c||}{-} & \multicolumn{1}{c}{1.12$\times 10^{15}$ } & \multicolumn{1}{c}{4.24$\times 10^{14}$} &\multicolumn{1}{c|}{3.44$\times 10^{15}$}& \multicolumn{1}{c}{4.86$\times 10^{17}$}& \multicolumn{1}{c}{-}\\[0.5ex]
%				Plane& \multicolumn{1}{c}{} & \multicolumn{1}{c}{}&\multicolumn{1}{c|}{} & \multicolumn{1}{c}{}&\multicolumn{1}{c}{} & \multicolumn{1}{c||}{} & \multicolumn{1}{c}{} & \multicolumn{1}{c}{} &\multicolumn{1}{c|}{}& \multicolumn{1}{c}{}& \multicolumn{1}{c}{}\\[0.5ex]
				$c$-axis & \multicolumn{1}{c}{1.534} & \multicolumn{1}{c}{1.48}&\multicolumn{1}{c|}{1.36} & \multicolumn{1}{c}{1.65$\pm$0.03} & \multicolumn{1}{c}{-}&\multicolumn{1}{c||}{-} & \multicolumn{1}{c}{2.63$\times 10^{15}$ } & \multicolumn{1}{c}{7.17$\times 10^{14}$} &\multicolumn{1}{c|}{3.11$\times 10^{15}$}& \multicolumn{1}{c}{9.51$\times 10^{17}$}& \multicolumn{1}{c}{-}\\[0.5ex]
				\multirow{2}{*}{Total} & \multirow{2}{*}{1.472} & \multirow{2}{*}{-} & \multicolumn{1}{c|}{\multirow{2}{*}{-}}&\multirow{2}{*}{-} & \multirow{2}{*}{1.44$\pm$0.15} &\multicolumn{1}{c||}{1.61}& \multirow{2}{*}{1.62$\times 10^{15}$ } & \multirow{2}{*}{-} &\multicolumn{1}{c|}{\multirow{2}{*}{-}}& \multirow{2}{*}{-}& \multicolumn{1}{c}{6.25$\times 10^{15}$}&\multicolumn{1}{c}{3.9$\times 10^{17}$}\\[0.5ex]
				& \multicolumn{1}{c}{} & \multicolumn{1}{c}{} & \multicolumn{1}{c|}{} & \multicolumn{1}{c}{}&\multicolumn{1}{c}{} &\multicolumn{1}{c||}{1.19}& \multicolumn{1}{c}{} & \multicolumn{1}{c}{} & \multicolumn{1}{c|}{}&\multicolumn{1}{c}{}& \multicolumn{1}{c}{$\pm$5.37$\times 10^{16}$}&\multicolumn{1}{c}{4.27$\times 10^{13}$}\\[0.5ex]
				\br 
			\end{tabular}
		}
		%\addtolength{\tabcolsep}{1pt}
		\label{dc_al}
		
	\end{center}
	
\end{table*}

In its most general form, the diffusion coefficient of an atom along any arbitrary axis, $x$,  is given by \cite{sd0}:
\begin{equation} 
D_x = \frac{1}{2}f\Gamma \langle x^2\rangle
\end{equation}
where $f$ and $\Gamma$ are the jump correlation factor and jump frequency, respectively. $\langle x^2\rangle$ is the mean square displacement along the axis, averaged over a large number of jumps. In the HCP lattice, an atom has two independent jumps to nearest neighbor (NN) vacant sites; (a) between adjacent basal planes (out-of-plane) and (b) within a basal plane.  A jump within a basal plane contributes to the diffusion only in $xy$-plane, while the jump out of the plane contributes to diffusion in both $xy-$plane and along $z-$axis as shown in Fig.~\ref{hcpfig}. Diffusivities along the $c-$axis (1D diffusion along $z-$ axis, $D_{\parallel}$) and in-plane (2D diffusion in $xy-$plane, $D_{\perp}$) are given as \cite{sd0, sd1, sd2}

\begin{eqnarray}
D_{\parallel}& = \frac{1}{2}f_{Az}\Gamma_{A} \frac{c^2}{4} \label{dp0}\\ 
D_{\perp} &= \frac{1}{4}(f_{Bx}\Gamma_{B} a^2+f_{Ax}\Gamma_{A} \frac{a^2}{3}) \label{dp1}
\end{eqnarray}
where $\Gamma_{A}$ and $\Gamma_{B}$ are the total jump frequencies between adjacent basal planes and within a basal plane, respectively. $f_{Ax}$ and $f_{Bx}$ are partial correlation factors for in-plane diffusion due to out-of-plane and in-plane jumps, respectively, and $f_{Az}$  is the partial correlation factor for the diffusion along the $c-$axis due to out-of-plane jumps.

For a vacancy, all the jump sequences are random and hence, all the correlation factors are equal to unity. Total jump frequency, $\Gamma = Zw$, where $Z$ is the number of NN sites into which a vacancy can jump to and $w$ is the frequency for individual jumps. Since an HCP lattice has 6 in-plane and 6 out-of-plane NN sites, $Z_{A} = Z_{B} = 6$. Substituting for $\Gamma$ in Eqs.~\ref{dp0} and \ref{dp1} give,
\begin{eqnarray}
D_{\parallel}& = \frac{3}{4}c^2w_{A} \label{p0}\\  
D_{\perp} &= \frac{1}{2}a^2(3w_{B} +w_{A})\label{p1}
\end{eqnarray}
where $w_{A,B}=D_0~\mathrm{exp}(-E_{A,B}/k_BT)$  are the frequencies for individual out-of-plane and in-plane jumps and $E_{A,B}$ are their activation barriers, $D_0$ is the pre-exponential factor and is taken as $2\times10^{13}s^{-1}$\cite{dft_barriers}, $T$ is the absolute temperature.  In all our present simulations we assumed that the pre-exponential factors are independent of temperature, which varies only 1.67 times from 300 K to 500 K (D$_0 = (k_BT/h)$exp$(S_m/k_B)$, where $S_m$ is the entropy of migration).
Activation barriers for the out-of-plane and in-plane jumps calculated using the MEAM potential are $E_{A} = 0.563$ eV and $E_{B} = 0.547$ eV, respectively. The total diffusivity ($D_{total}$) of a vacancy is given as,
\begin{equation}
D_{total} = \frac{1}{3}(2D_{\perp}+ D_{\parallel}) \label{dtotal}
\end{equation}

Diffusion coefficient (D) of a randomly diffusing entity, which in our case is a vacancy, is obtained using the Einstein equation:\cite{msd} 
\begin{equation}
D = \frac{1}{2d}lim_{t\rightarrow\infty}\frac{\langle \Delta r(t)^2\rangle}{t} = \frac{1}{2d}\frac{MSD}{t}
\end{equation}
where $\Delta r(t)$ and $\langle \Delta r(t)^2\rangle$ are the position and mean square displacement (MSD) at time $t$, respectively and $d$ is the dimensionality of the diffusion. $MSD/t$ is obtained from the slope of MSD Vs.~time plot. Note that the dimensionality of vacancy diffusion within and perpendicular to the basal plane are 2 and 1, respectively. 
Table.~\ref{dc} shows a comparison of vacancy diffusivities in pure Mg obtained from KMC simulations and analytical Eqs.~\ref{p0}-\ref{dtotal} at temperatures of 300, 400 and 500 K. Good agreement between KMC simulation and analytical values show that the KMC code is working correctly and the HCP lattice is mapped correctly on to the simple cubic lattice.

\begin{figure}  
	\centering
	\subfloat[Vacancy and Al atom within the basal plane (A-layer)]{
		\includegraphics[width=1.95in]{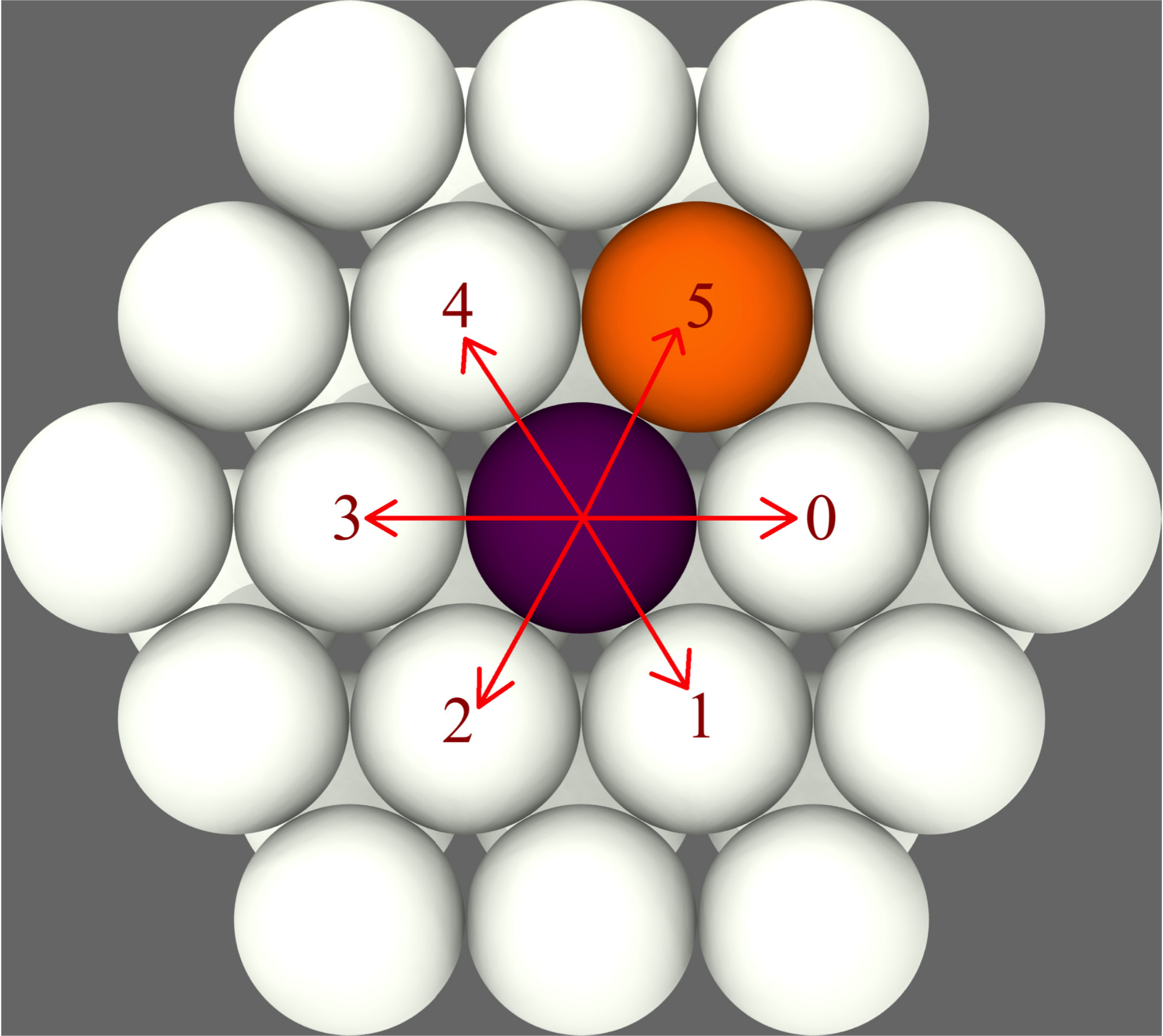}
		\label{fig6a}
	}\\
	\subfloat[Al atoms at NN sites within the basal plane (B-layer)]{
		\includegraphics[width=1.95in]{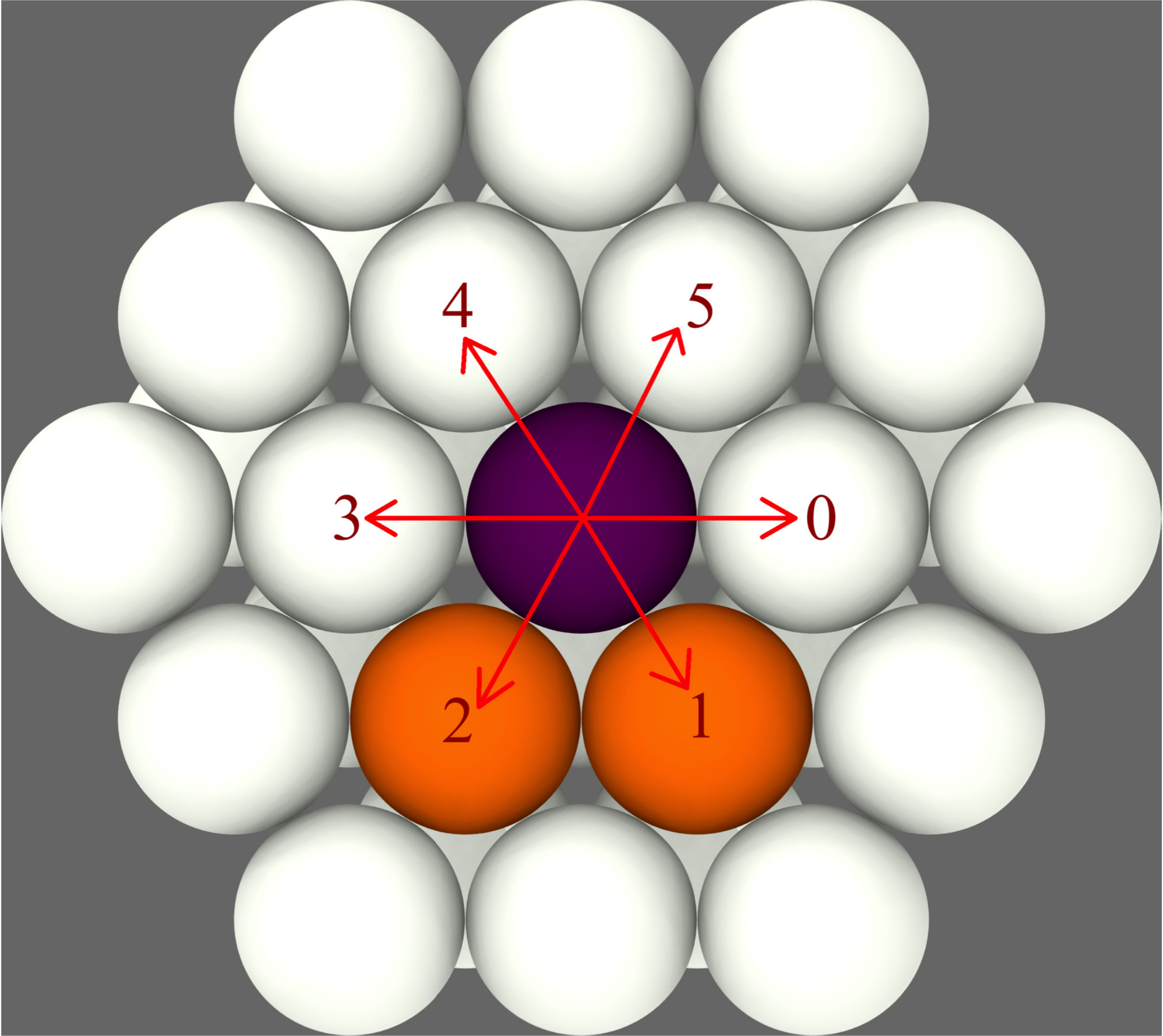}  
		\label{fig6b}
	}\\
	\subfloat[Al atom at diagonally opposite NN sites within the basal plane (A-layer)]{
		\includegraphics[width=1.95in]{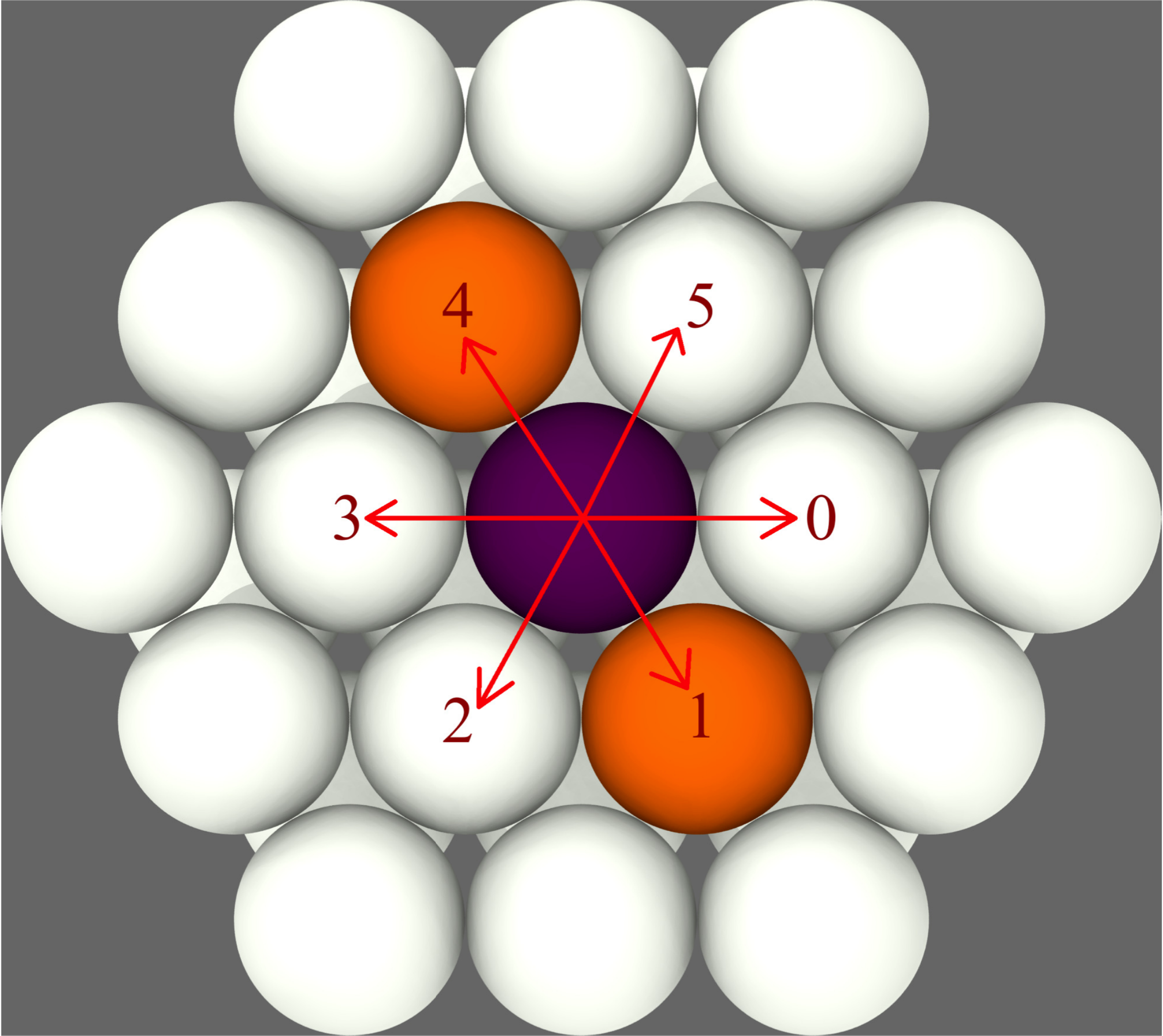}  
		\label{fig6c}
	}\\
	\subfloat[Vacancy and Al atom in adjacent planes (A-layer)]{
		\includegraphics[width=1.95in]{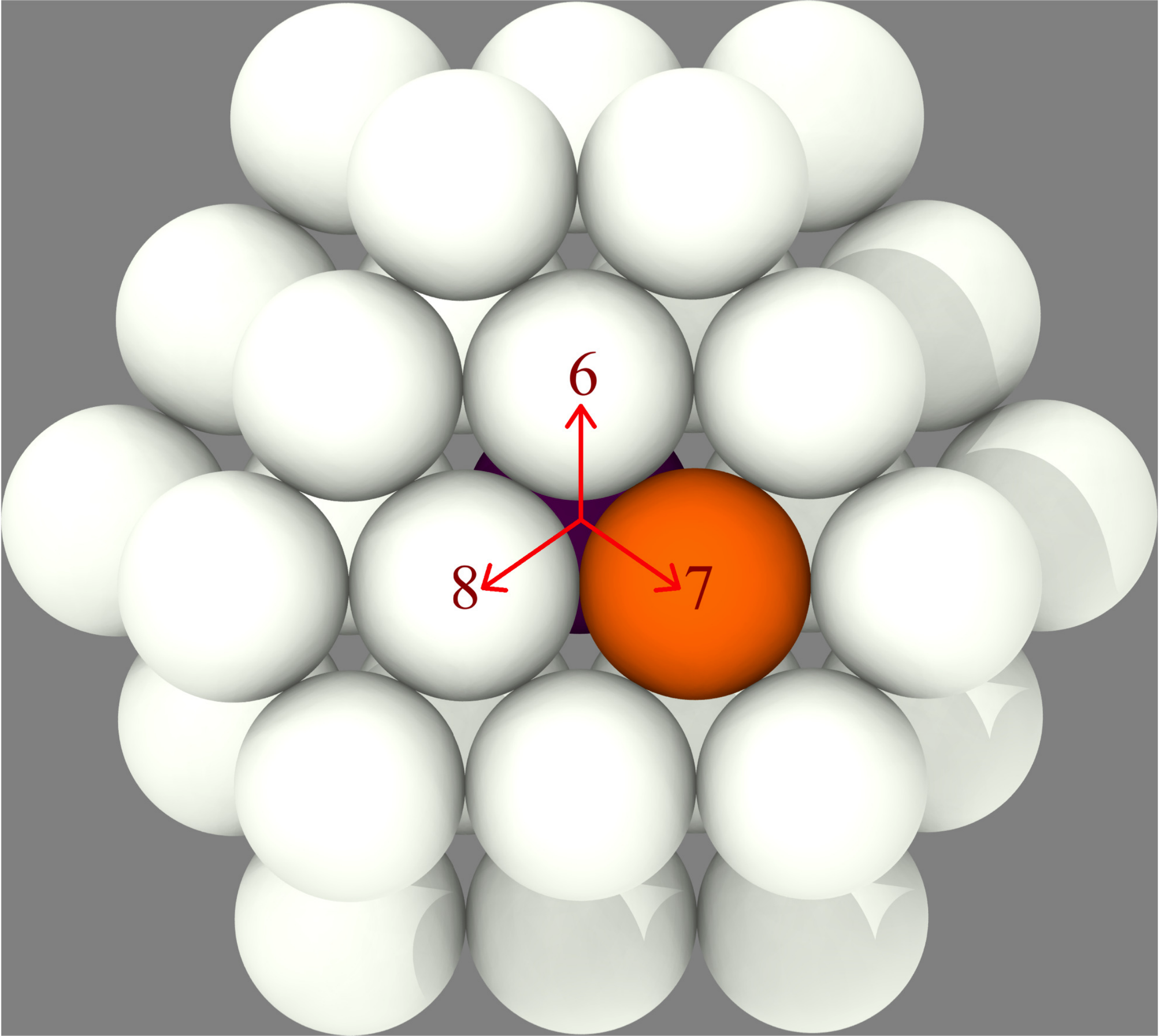}  
		\label{fig6d}
	}
	
	\caption{Al atom neighborhoods around a vacancy. White, purple and orange spheres represent Mg atoms, vacancy and Al atoms, respectively.}
	\label{fig6}  
\end{figure}

\begin{figure*}
	\centering
	\includegraphics[width=6.00in]{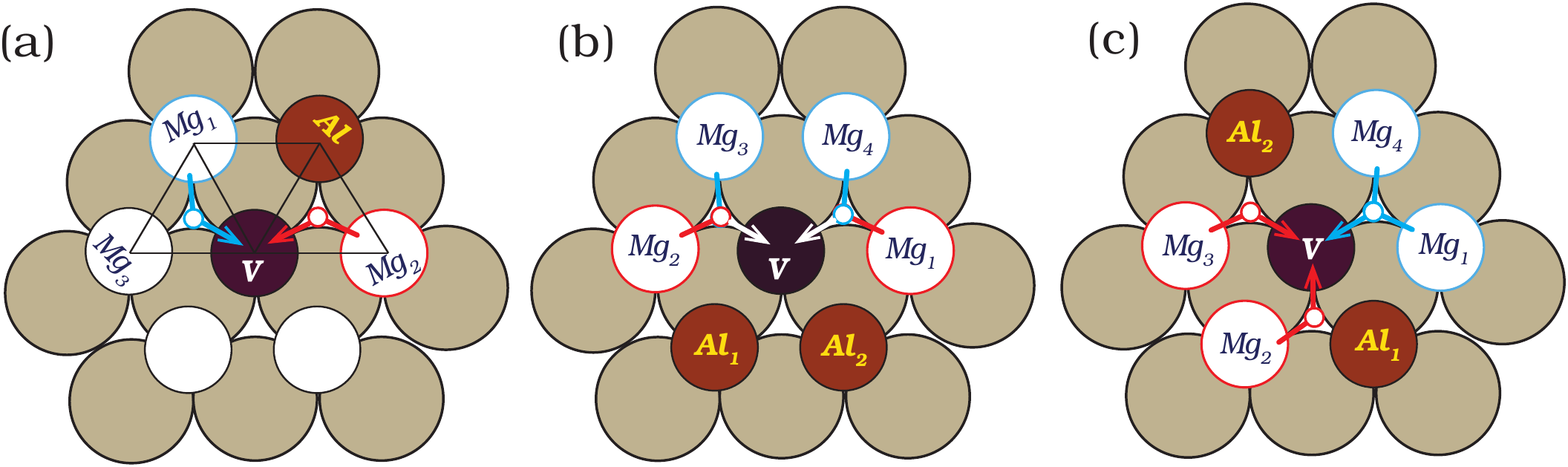}
	\caption{(a) Top view of Fig.~\ref{fig6a}. Blue and red arrows correspond to P$_0$ and P$_4$, respectively  (b) Top view of Fig.~\ref{fig6b}. P$_0$ \& P$_3$ (red arrows) and P$_4$ \& P$_5$ (blue arrows) are pairs of symmetric hops  (c) Top view of Fig.~\ref{fig6c}.P$_0$ \& P$_5$ (blue arrows) and P$_2$ \& P$_3$ (red arrows) are pairs of symmetric hops. Arrows show (approximately) the path of Mg atom and the small circle in middle represents the saddle point. (for the sake of clarity, atoms on the top layer are shown by the smaller circles)}
	\label{figa}
\end{figure*}

\subsection{Al Atom Diffusion in Mg}

Vacancy mediated solute diffusivity is given as
\begin{eqnarray}
D_{s} =a^2f_2C_V(s)\Gamma_{2} \label{Dii} 
\end{eqnarray}
where $f_2$ is the solute correlation factor, $\Gamma_2$ is the vacancy-solute exchange frequency, $C_V (s)  = C_V^{eq} \mathrm{exp}(-E^{V-s}_{b}/k_BT)$, is the vacancy concentration near a solute atom and $E_b^{V-s}$ is the vacancy-solute binding energy. In the case of solute diffusion, $f_2$ is not a geometric factor, but depends on the atomic jump frequencies of the vacancies and the temperature. Binding energies ($E_b^{V-s}$) calculated using the MEAM potential, when the vacancy and Al atom are at NN locations that are either in the same plane or in the adjacent planes, are very small (-0.01 eV$\not<E_b^{V-s}\not>$ 0.005 eV) and is in agreement with a recent DFT study.\cite{al_dft_new} Therefore, in our simulations $E_b^{V-s}$ is taken as zero. The Eq.~\ref{Dii} can be written as,
\begin{eqnarray}
\label{Diif}
D_{s} = D_0a^2f_2\mathrm{exp}\Bigg(\frac{\Delta S^V_f}{k_B}\Bigg)\mathrm{exp}\Bigg(-\frac{E^V_f+E_{x}}{k_BT}\Bigg)
\end{eqnarray}
where $E_{x}$ is the effective activation barrier for vacancy and Al atom exchange.
For SLKMC simulations of an Al atom diffusion, the simulation cell has one vacancy and one Al atom. Note that to calculate Al atom diffusivities, the actual physical time is obtained by rescaling the KMC simulation time according to Eq.~\ref{treal}.

Table.~\ref{dc_al} shows a comparison of effective activation energies and effective diffusion prefactors for an Al atom diffusion (within and perpendicular to the basal plane, and total diffusion) in Mg matrix  obtained from SLKMC simulations, 8-frequency model \cite{Al_diff_dft, al_dft_new}, diffusion couple experiment using single crystal Mg \mcite{*all_diff_exp1, *all_diff_exp11},  solid-to-solid diffusion couple of polycrystalline Mg and Mg-Al solid solution \cite{all_diff_exp2, Brennen2} and via depth profiling of Al penetration in polycrystalline Mg ($\sim 10~\mu m$ grain size).\cite{brennan} To obtain Al atom diffusivities, Das {\it et al.} \mcite{*all_diff_exp1, *all_diff_exp11}, extrapolated interdiffusion coefficients in hcp Mg-Al solid solution to 0\%  Al concentration using Wagner's approach.\cite{wagner} While Kammerer {\it et al.} \cite{all_diff_exp2} ($\sim100-500~\mu m$ grain size) and Brennan {\it et al.} \cite{Brennen2} ($\sim 30-60~\mu m$ grain size) extrapolated interdiffusion coefficients  obtained using the Boltzmann-Matano method \mcite{*BM1, *BM2} to less than 1at.\% Al concentration of Mg-Al solid solution using the Hall method \cite{Hall}. Brennan {\it et al}\cite{brennan} used depth profiling with secondary ion mass spectrometry (SIMS) and, utilized the thin film method and the diffusion equation for the thin film solution to extract the Al atom diffusivities. 
Ganeshan {\it et al.} \cite{Al_diff_dft} and Zhou {\it et al.}\cite{al_dft_new} used 8-frequency model\cite{8freq} coupled with DFT-LDA and DFT-PBEsol calculated vacancy-atom (Mg \& Al) exchange frequencies, respectively, to obtain Al atom diffusivities in a single crystal Mg. Note that the Al atom diffusivities obtained in present simulations are also for a single crystal or a polycrystalline Mg with a very large grain size.

\begin{table*}
	\footnotesize
	\caption{Activation barriers for vacancy hops to 12 NN in the presence of an Al atom. P$_0$ to P$_{11}$ represent vacancy hops in directions 0 to 11 as shown in Fig.~\ref{fig6}. Basis of 0 and 1 represent whether the vacancy is in an A- or B-layer}
	
	\begin{center}
		\begin{tabular}{@{}lc|cccccc|ccc|ccc}
			\br
			\multirow{3}{*}{Pattern} && \multicolumn{6}{c|}{within basal plane}&\multicolumn{3}{c|}{To the plane above}&\multicolumn{3}{c}{To the plane below}\\[0.5ex]
			\cline{2-14}
			&&&&&&&&&&&\\
			\ns
			&Basis& P$_0$ & P$_1$ & P$_2$ & P$_3$ & P$_4$ &P$_5$ &P$_6$&P$_7$&P$_8$&P$_9$&P$_{10}$ & P$_{11}$\\[0.5ex]
			\hline
			&&&&&&&&&&&\\
			\ns
			Fig~\ref{fig6a}& 0 & 0.283 & 0.538 & 0.533 & 0.533 & 0.543 & 0.574 & 0.398& 0.643 & 0.554 & 0.398 & 0.643 & 0.554 \\
			Fig~\ref{fig6b}& 1 & 0.313 & 0.646 & 0.646 & 0.313 & 0.464 & 0.464& 0.154& 0.644 & 0.644 & 0.154 & 0.644 & 0.644 \\
			Fig~\ref{fig6c}& 0 & 0.520 & 0.601 & 0.275 & 0.275 & 0.601 & 0.520 & 0.391& 0.392 & 0.729 & 0.392 & 0.391 & 0.729 \\
			Fig~\ref{fig6d}& 0 & 0.424 & 0.424 & 0.627 & 0.546 & 0.546 & 0.627 & 0.414& 0.637 & 0.414 & 0.550 & 0.588 & 0.549 \\
			\br
		\end{tabular}
	\end{center}
	\label{proc}
\end{table*}

Al atom diffusivities obtained from SLKMC simulations using the MEAM potential are in qualitative agreement with the published experimental and 8-frequency model values.  Our simulations also show that an Al atom diffuses faster within the basal plane than along the $c$-axis, 
The effective activation barriers obtained from SLKMC simulations are a tenth of an eV higher when compared to the values obtained by Ganeshan {\it et al.} \cite{Al_diff_dft} and Zhou {\it et al}\cite{al_dft_new}. While the effective prefactors have the same order of magnitude as those obtained by Zhou {\it et al.}; the values obtained by Ganeshan {\it et al.} are an order of magnitude lower. Note that in a recent work Allnat {\it et al.}\cite {13_freq} show that to fully consider the anisotropy of an HCP lattice, 13 independent atom-vacancy exchange frequencies are needed in contrast to 8 needed for the 8-frequency model. Also, no closed form expressions to calculate correlation factors exist for the 13-frequency model. In SLKMC simulation, the anisotropy of an HCP lattice is naturally included as well as all the possible vacancy-Mg and vacancy-Al exchange processes. In comparison to those obtained by  Kammerer {\it et al.} \cite{all_diff_exp2}, the activation energy for the total diffusion is within the error bars and the effective prefactor has the same order of magnitude, but the effective prefactor has a very large error bar. On the other hand, prefactors obtained from diffusion couple experiments with single crystal Mg by Das {\it et al.} \mcite{*all_diff_exp1,*all_diff_exp11} are  2-to-3 orders of magnitude higher than those obtained from 8-frequency model and SLKMC simulations. Surprisingly, both activation energies and prefactors obtained by Das {\it et al.} \mcite{*all_diff_exp1,*all_diff_exp11} are in good agreement  with those obtained by Brennan {\it et al.} \cite{brennan}  using a polycrystalline Mg with a grain size of 10 $\mu$m. In another study Brennan {\it et al.} \cite{Brennen2}, using a polycrystalline Mg with a grain size of 30 to 60 $\mu$m, obtained a  prefactor which is 4 orders of magnitude lower than their previous value.
%In a similar experiment by Brennan {\it et al.} using a polycrystalline larger grain size ($\sim 30-60~\mu m$) obtained an effect \cite{brennan} 
Based on the available data, although the activation barriers differ only by few tenths of an eV, there seems to be a clear disagreement in the values of prefactors obtained using experiments and theoretical values.

\begin{table*}
	\caption{Activation barriers for vacancy hops to 12 NN for Al-atom neighborhoods shown in Fig.~\ref{patterns}. Basis of 0 and 1 represent whether the vacancy is in A- or B-layer}
	\begin{center}
		\footnotesize
		\begin{tabular}{@{}lc|cccccc|ccc|ccc}
			\br
			\multirow{3}{*}{Pattern} && \multicolumn{6}{c|}{Within basal plane} &\multicolumn{3}{c|}{To the plane above}&\multicolumn{3}{c}{To the plane above}\\[0.5ex]
			\cline{2-14}
			&&&&&&&&&&&\\
			\ns
			&Basis& P$_0$ & P$_1$ & P$_2$ & P$_3$ & P$_4$ &P$_5$ &P$_6$&P$_7$&P$_8$&P$_9$&P$_{10}$ & P$_{11}$\\[0.5ex]
			\hline
			&&&&&&&&&&&\\
			\ns
			Fig~\ref{fig7a}& 1 & 0.535 & 0.547 & 0.550 & 0.566 & 0.552 & 0.550 & 0.555& 0.564 & 0.578 & 0.574 & 0.547 & 0.560 \\
			Fig~\ref{fig7b}& 0 & 0.535 & 0.551 & 0.552 & 0.556 & 0.550 & 0.547& 0.574& 0.560 & 0.547 & 0.555 & 0.578 & 0.564 \\
			Fig~\ref{fig7c}& 1 & 0.627 & 0.423 & 0.423 & 0.627 & 0.545 & 0.545 & 0.588& 0.550 & 0.550 & 0.640 & 0.413 & 0.413 \\
			Fig~\ref{fig7d}& 0 & 0.423 & 0.424 & 0.627 & 0.546 & 0.544 & 0.627 & 0.414& 0.637 & 0.414 & 0.550 & 0.588 & 0.549 \\
			Fig~\ref{fig7e}& 0 & 0.283 & 0.538 & 0.533 & 0.533 & 0.543 & 0.573 & 0.398& 0.645 & 0.554 & 0.398 & 0.644 & 0.552 \\
			Fig~\ref{fig7f}& 1 & 0.573 & 0.283 & 0.539 & 0.533 & 0.534 & 0.543 & 0.644& 0.556 & 0.397 & 0.644& 0.556 & 0.397 \\
			Fig~\ref{fig7g}& 0 & 0.554& 0.553 & 0.545 & 0.552 & 0.552 & 0.544 & 0.562& 0.564 & 0.561 & 0.562& 0.564 & 0.561 \\
			Fig~\ref{fig7h}& 1 & 0.552& 0.552 & 0.554 & 0.554 & 0.553 & 0.545 & 0.562& 0.564 & 0.561 & 0.562& 0.564 & 0.561 \\
			Fig~\ref{fig7i}& 1 & 0.560& 0.533 & 0.611 & 0.611 & 0.553 & 0.556 & 0.554& 0.553 & 0.562 & 0.562& 0.644 & 0.552 \\
			Fig~\ref{fig7j}& 0 & 0.611& 0.533 & 0.556 & 0.557 & 0.533 & 0.611 & 0.554& 0.553 & 0.562 & 0.562& 0.644 & 0.552 \\
			Fig~\ref{fig7k}& 1 & 0.542& 0.545 & 0.549 & 0.523 & 0.548 & 0.565 & 0.565& 0.579 & 0.544 & 0.565& 0.579 & 0.544 \\
			Fig~\ref{fig7l}& 0 & 0.550& 0.523 & 0.548 & 0.542 & 0.545 & 0.549 & 0.565& 0.579 & 0.544 & 0.565& 0.579 & 0.544 \\
			Fig~\ref{fig7m}& 1 & 0.560& 0.528 & 0.553 & 0.554 & 0.528 & 0.560 & 0.579& 0.578 & 0.572 & 0.579& 0.578 & 0.572 \\
			Fig~\ref{fig7n}& 0 & 0.554& 0.528 & 0.559 & 0.560 & 0.528 & 0.553 & 0.578& 0.572 & 0.579 & 0.579& 0.578 & 0.572 \\
			Fig~\ref{fig7o}& 0 & 0.550& 0.561 & 0.537 & 0.548 & 0.558 & 0.552 & 0.564& 0.563 & 0.569 & 0.568& 0.552 & 0.547 \\
			Fig~\ref{fig7p}& 1 & 0.559& 0.547 & 0.537 & 0.562 & 0.549 & 0.552 & 0.570& 0.563 & 0.563 & 0.549& 0.552 & 0.567 \\
			Fig~\ref{fig7q}& 0 & 0.548& 0.548 & 0.548 & 0.548 & 0.548 & 0.548 & 0.588& 0.587 & 0.587 & 0.573& 0.572 & 0.572 \\
			Fig~\ref{fig7r}& 1 & 0.548& 0.548 & 0.548 & 0.548 & 0.548 & 0.548 & 0.588& 0.587 & 0.587 & 0.573& 0.572 & 0.572 \\
			\br
		\end{tabular}
	\end{center}
	\label{All}
\end{table*}

\section{Examples of Vacancy Hops in the Presence of Al Atoms}\label{AlMgx}

Here we show that the individual jump frequencies of a vacancy in Mg matrix, within and out of a basal plane ($w_{A, B}$) in various directions, and are no longer equal each other in the presence of an Al atom.
Fig.~\ref{fig6}  shows examples for vacancy-Mg and vacancy-Al exchange processes and the corresponding activation barriers are given in Table.~\ref{proc}. Figs.~\ref{fig6}(a-c) show vacancy jumps to NN sites within the basal plane and Fig.~\ref{fig6d} shows vacancy jumps to the three NN sites in the adjacent plane above. Note that the directions shown in Fig.~\ref{fig6d} are for a vacancy in an A-layer, but  the jump directions for a vacancy in a B-layer are equivalent and can be obtained by rotating the directions  shown in  Fig.~\ref{fig6d} by $120^{\circ}$ either clockwise or anti-clockwise. Also, note that P$_0$ to P$_{11}$  in Table.~\ref{proc} represent vacancy hops in directions 0 to 11 shown in Fig.~\ref{fig6}.

Vacancy-Al atom exchanges always have the largest activations barriers  when compared to vacancy-Mg atom exchange processes, irrespective of whether they are located at NN sites that are within the basal plane or in adjacent basal planes (P$_5$ in Figs.~\ref{fig6a} and P$_7$ in Fig.~\ref{fig6d}). The activation barrier when vacancy and Al are at NN sites in adjacent planes is always the larger one. Accordingly, vacancy and Al atom exchange occurs more often within the basal plane and hence, Al atom diffuses faster perpendicular to the $c-$axis. The activation barrier for a vacancy and the Mg-atom which is NN both to the vacancy and the Al atom has the lowest activation barrier (P$_0$ in Fig.~\ref{fig6a}), while a similar exchange between adjacent planes has the second lowest activation barrier (P$_6$ and P$_9$ in Fig.~\ref{fig6a}). Although the activation barriers are different, qualitatively similar behavior was observed even in DFT calculated activation barriers for the same processes. \cite{dft_barriers} From Figs.~\ref{fig6}(b-c) and Table.~\ref{proc} one can see that the activation barriers for the vacancy-Al atom exchange increases while for the vacancy-Mg atom exchange  decreases with increasing number of Al atoms. When the Al atoms are at NN sites on the same plane, the activation barrier for the exchange of vacancy and Mg atom located on adjacent planes, but are NN to the Al atoms, is the lowest (P$_6$ and P$_9$ in Fig~\ref{fig6b}). The activation barrier for vacancy-Al atom exchange is even larger when the two Al atoms are located at NN sites on adjacent planes (not shown). This suggests that a single vacancy will diffuse faster with increasing Al concentration, but the Al diffusivity will decrease due an increase in the activation barrier for vacancy-Al atom exchange.

\begin{figure*}  
	\centering
	\subfloat[ 1]{
		\includegraphics[width=0.75in]{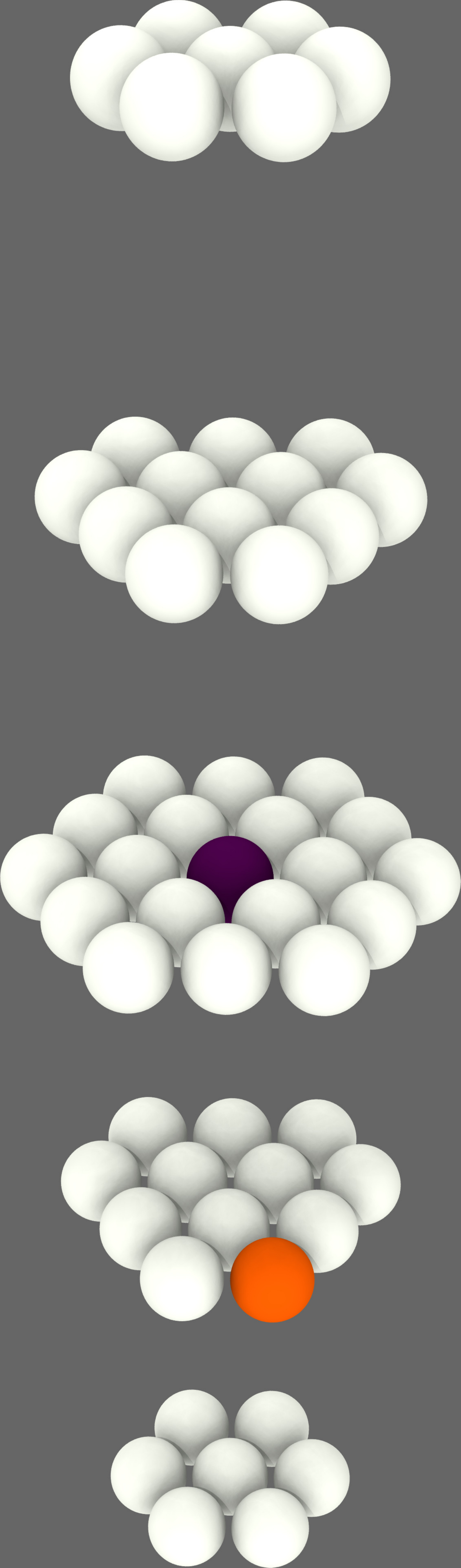}
		\label{fig7a}
	}
	\subfloat[ 2]{
		\includegraphics[width=0.75in]{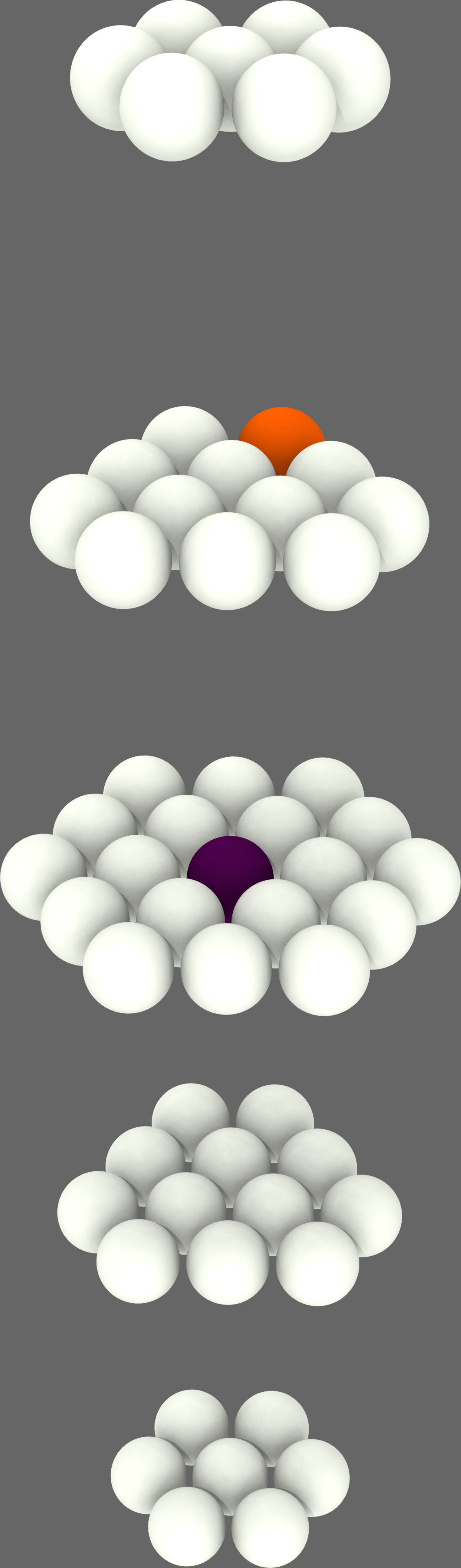}  
		\label{fig7b}
	}
	\subfloat[ 3]{
		\includegraphics[width=0.75in]{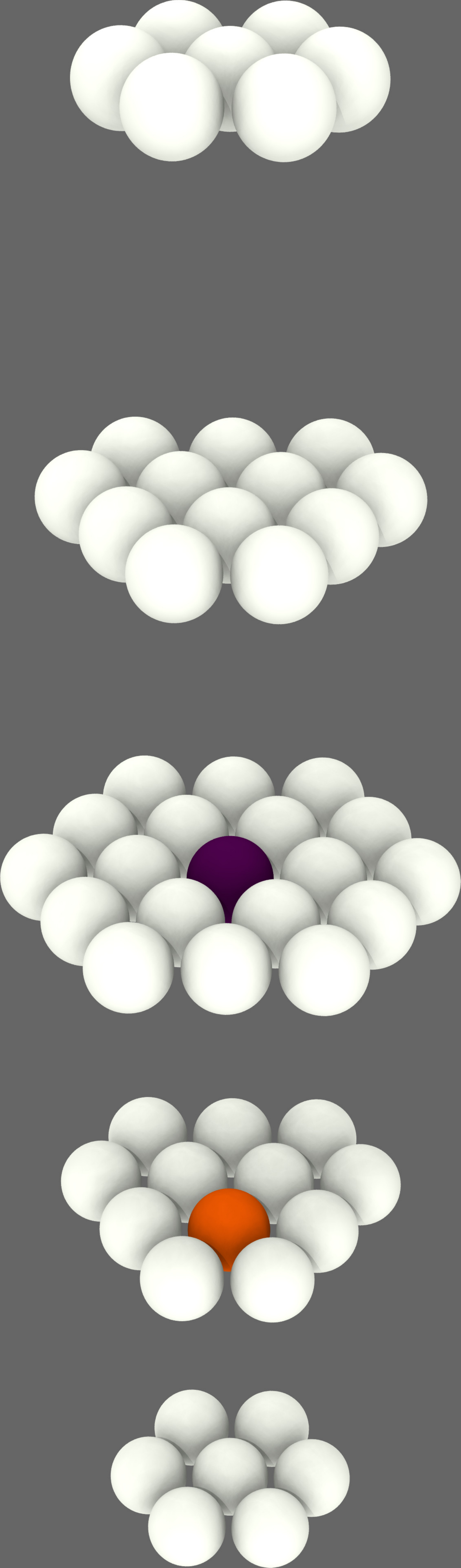}  
		\label{fig7c}
	}
	\subfloat[ 4]{
		\includegraphics[width=0.75in]{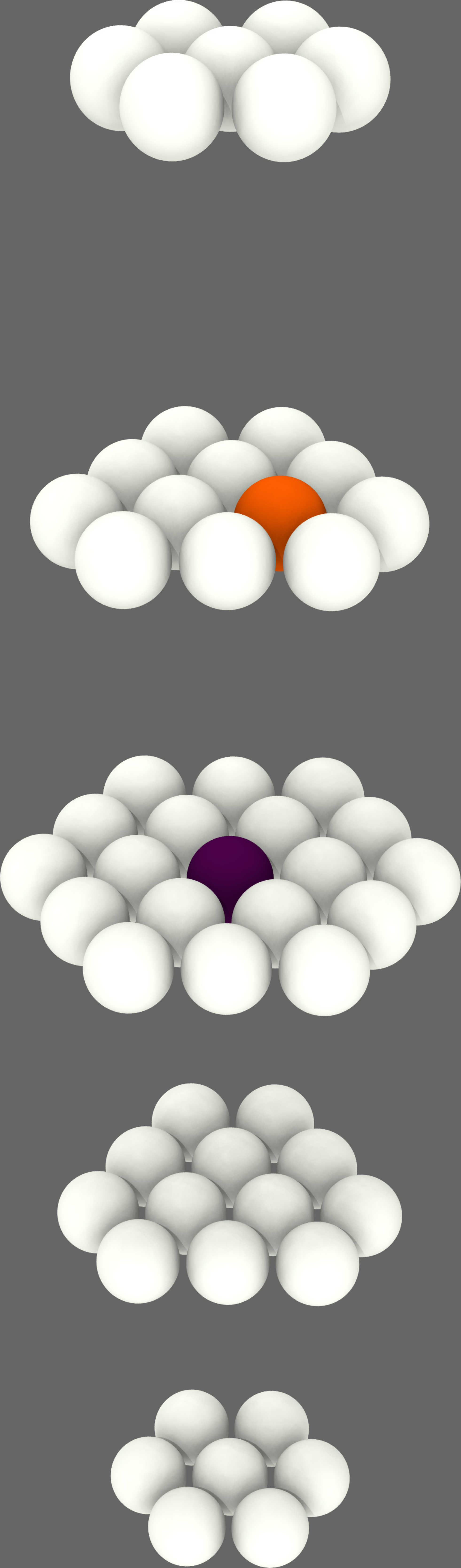}  
		\label{fig7d}
	}
	\subfloat[ 5]{
		\includegraphics[width=0.75in]{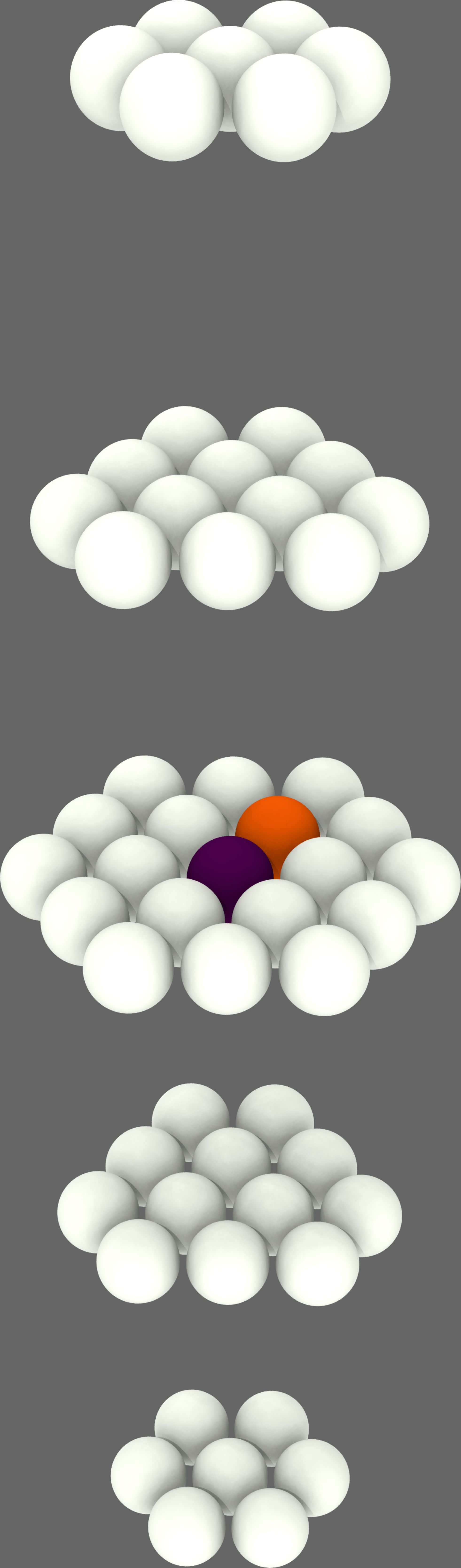}  
		\label{fig7e}
	}
	\subfloat[ 6]{
		\includegraphics[width=0.75in]{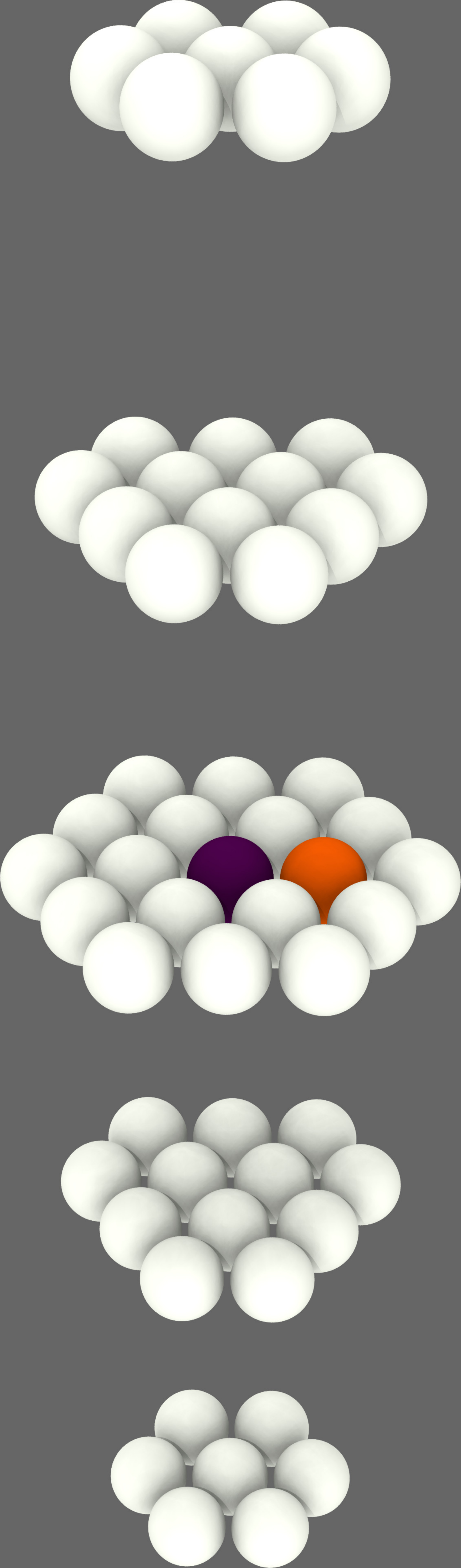}  
		\label{fig7f}
	}\\
	\subfloat[ 7]{
		\includegraphics[width=0.75in]{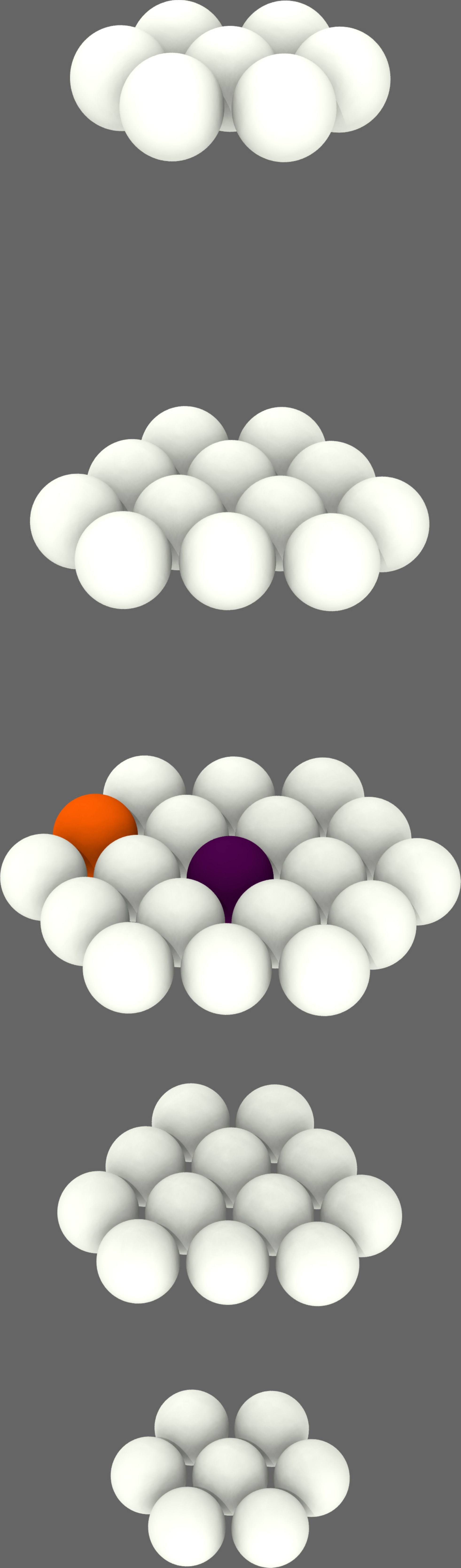}  
		\label{fig7g}
	}
	\subfloat[ 8]{
		\includegraphics[width=0.75in]{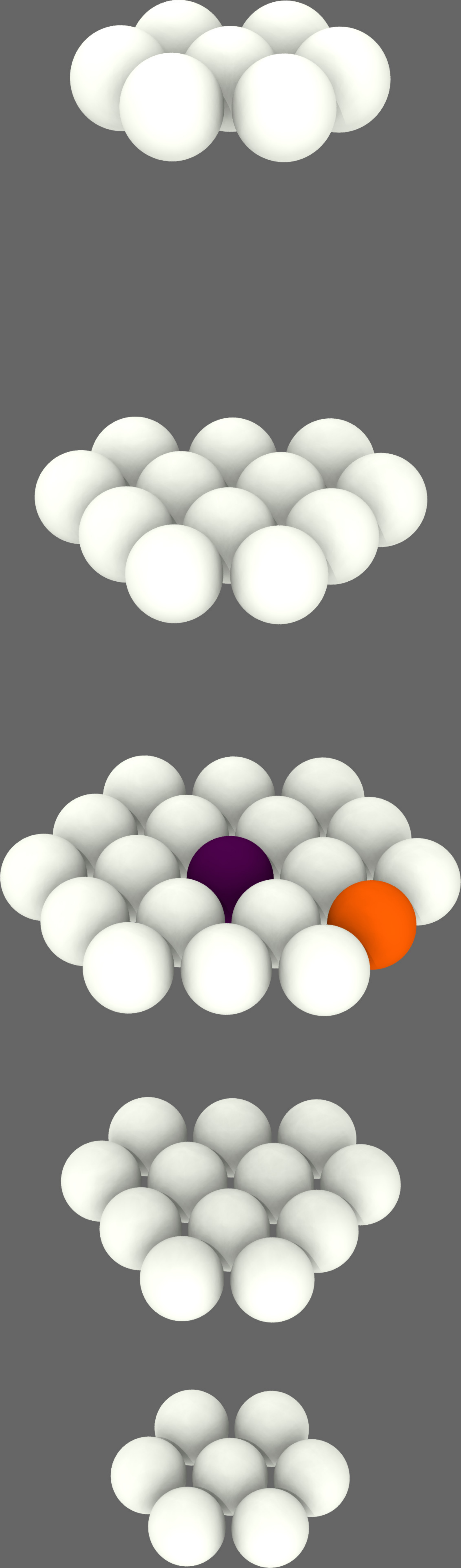}  
		\label{fig7h}
	}
	\subfloat[ 9]{
		\includegraphics[width=0.75in]{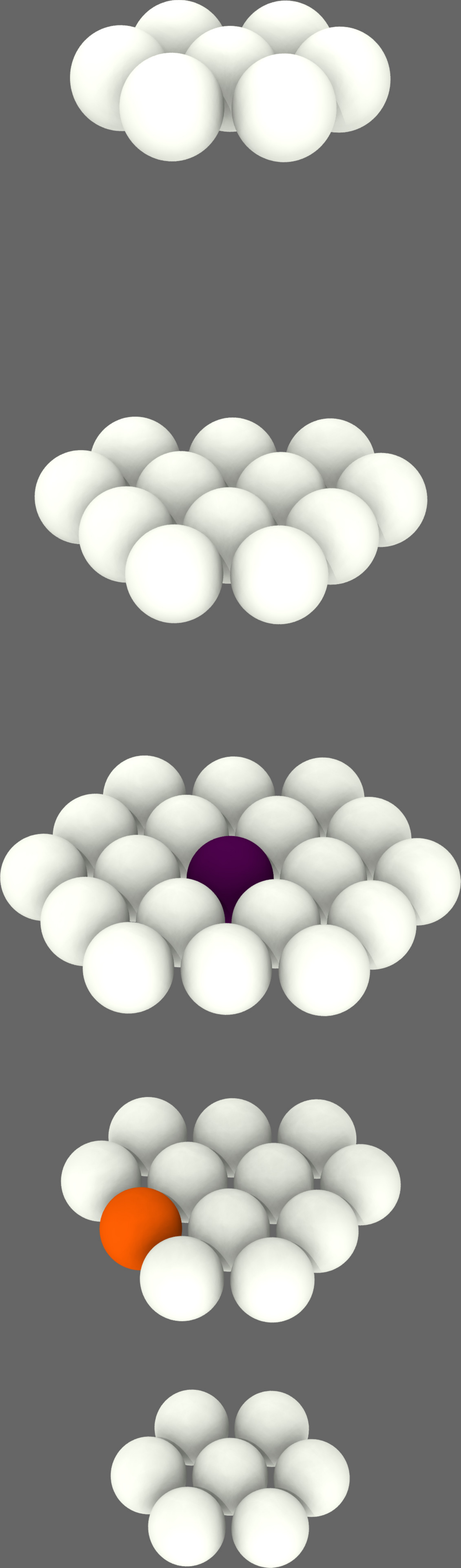}  
		\label{fig7i}
	}
	\subfloat[ 10]{
		\includegraphics[width=0.75in]{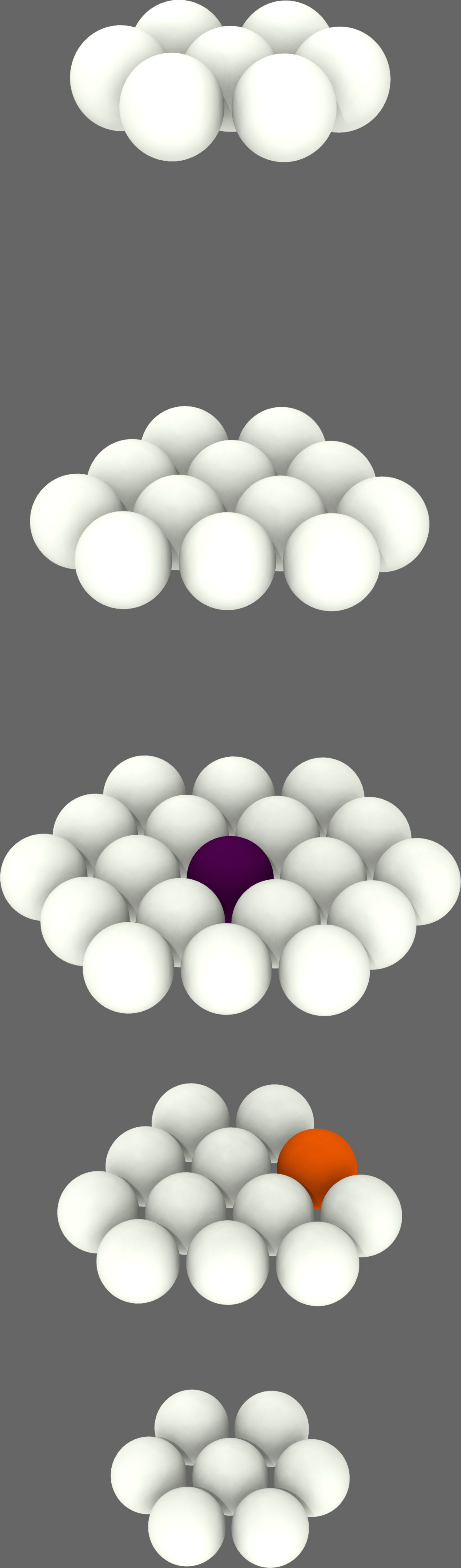}  
		\label{fig7j}
	}
	\subfloat[ 11]{
		\includegraphics[width=0.75in]{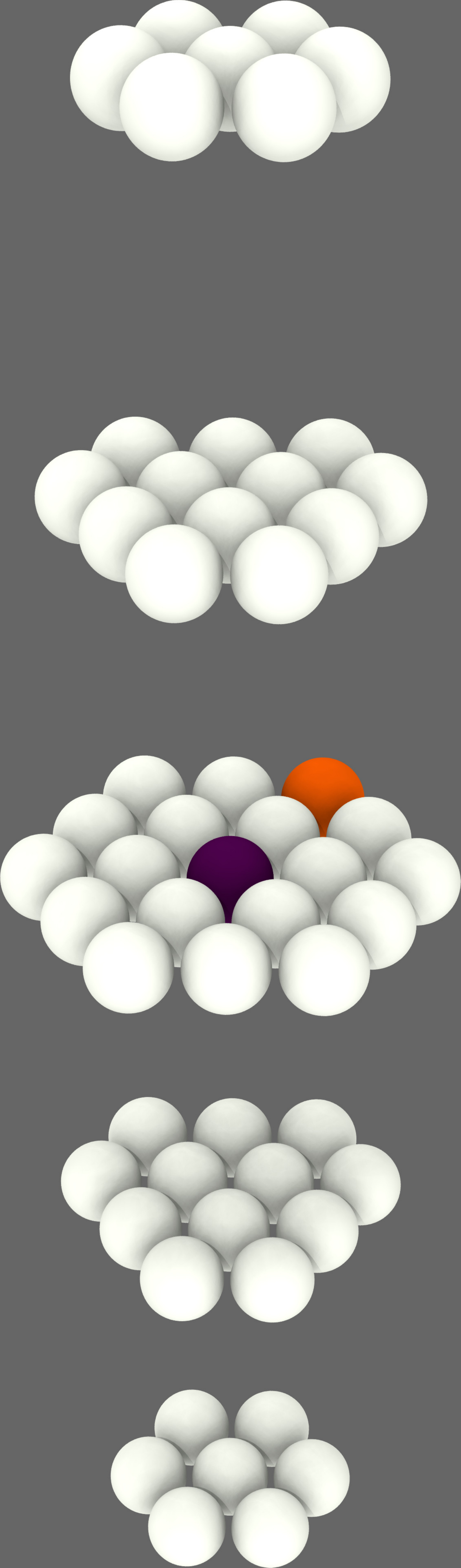}  
		\label{fig7k}
	}
	\subfloat[ 12]{
		\includegraphics[width=0.75in]{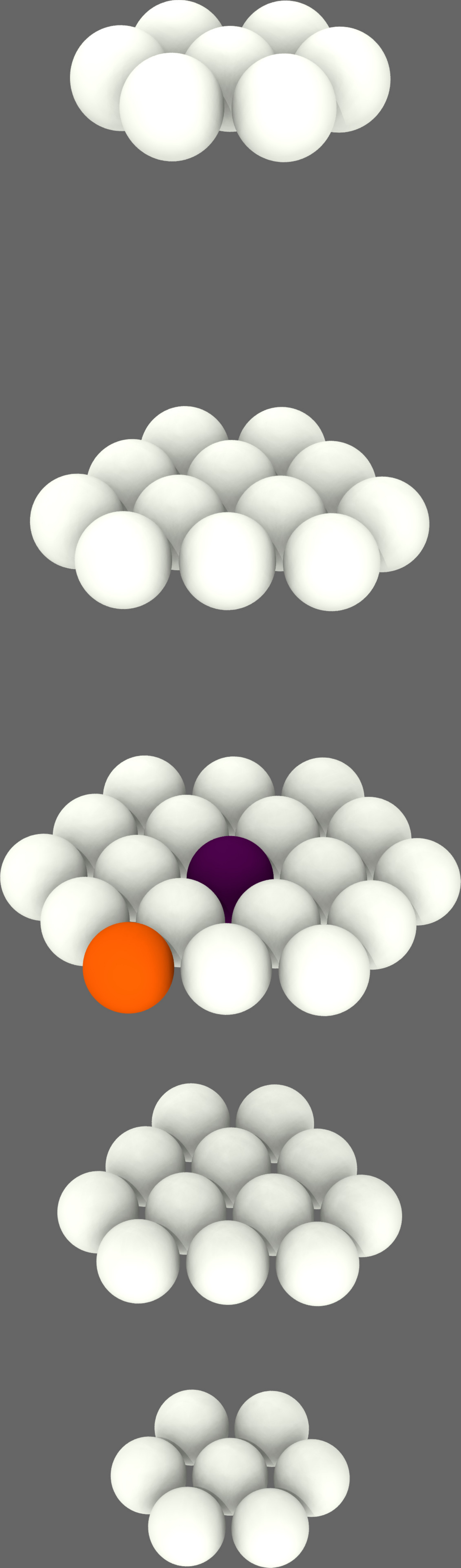}  
		\label{fig7l}
	}\\
	\subfloat[ 13]{
		\includegraphics[width=0.75in]{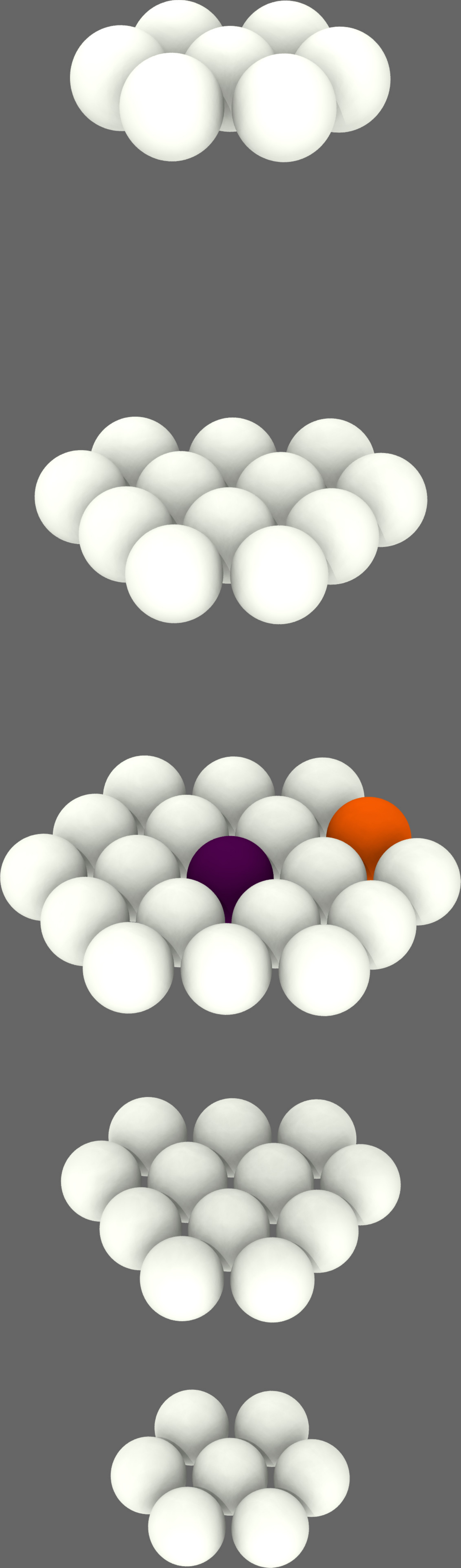}  
		\label{fig7m}
	}
	\subfloat[ 14]{
		\includegraphics[width=0.75in]{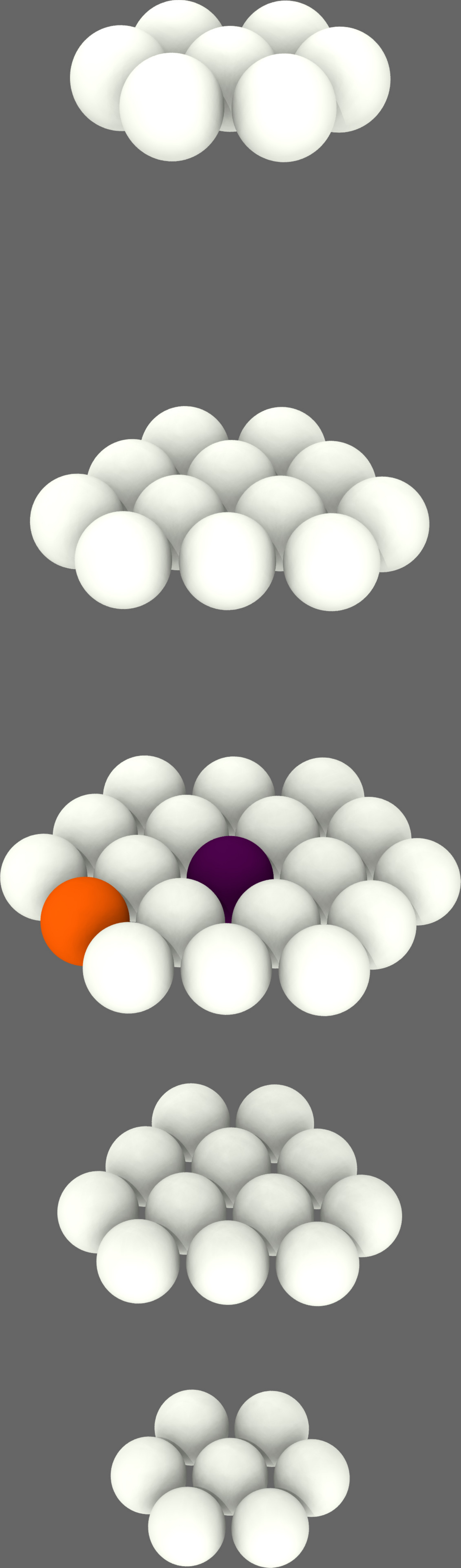}  
		\label{fig7n}
	}
	\subfloat[ 15]{
		\includegraphics[width=0.75in]{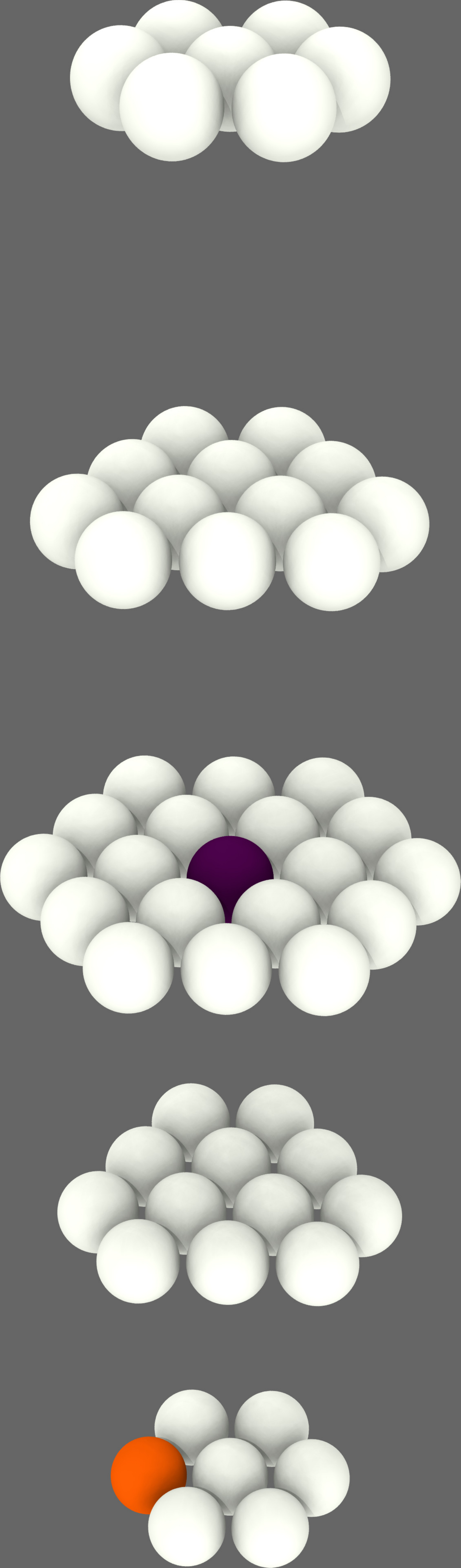}  
		\label{fig7o}
	}
	\subfloat[ 16]{
		\includegraphics[width=0.75in]{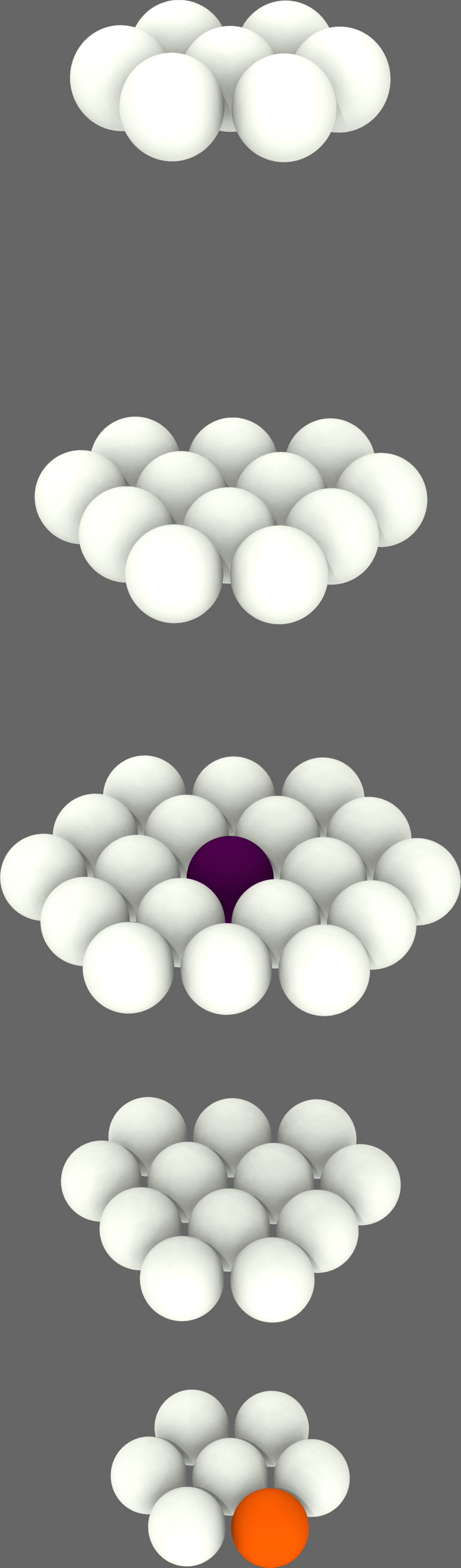}  
		\label{fig7p}
	}
	\subfloat[ 17]{
		\includegraphics[width=0.75in]{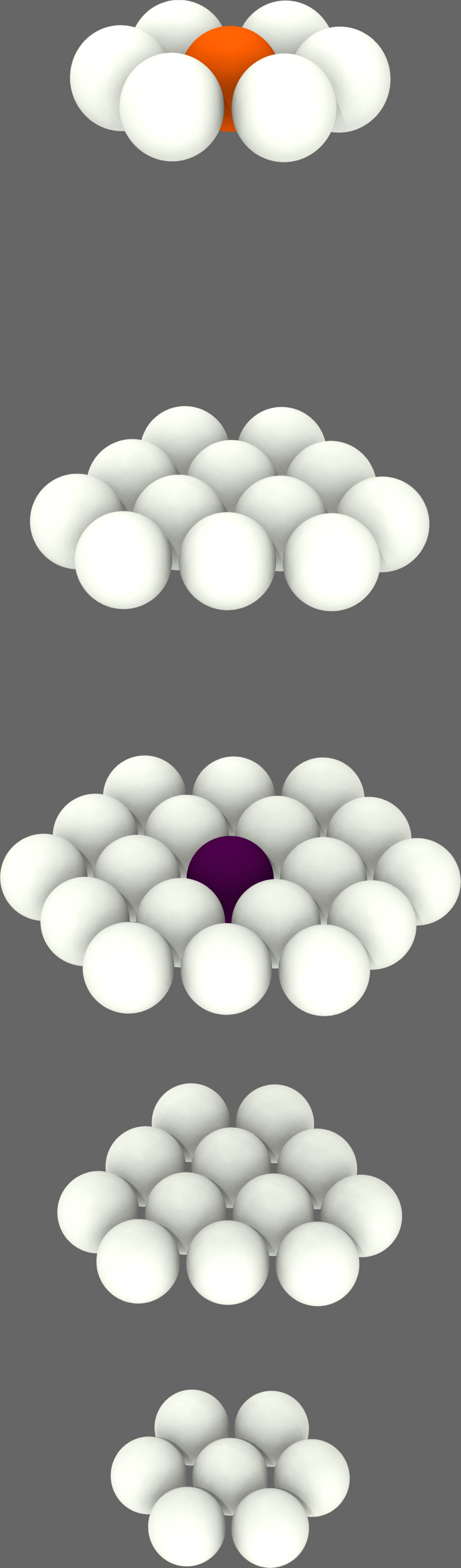}  
		\label{fig7q}
	}
	\subfloat[ 18]{
		\includegraphics[width=0.75in]{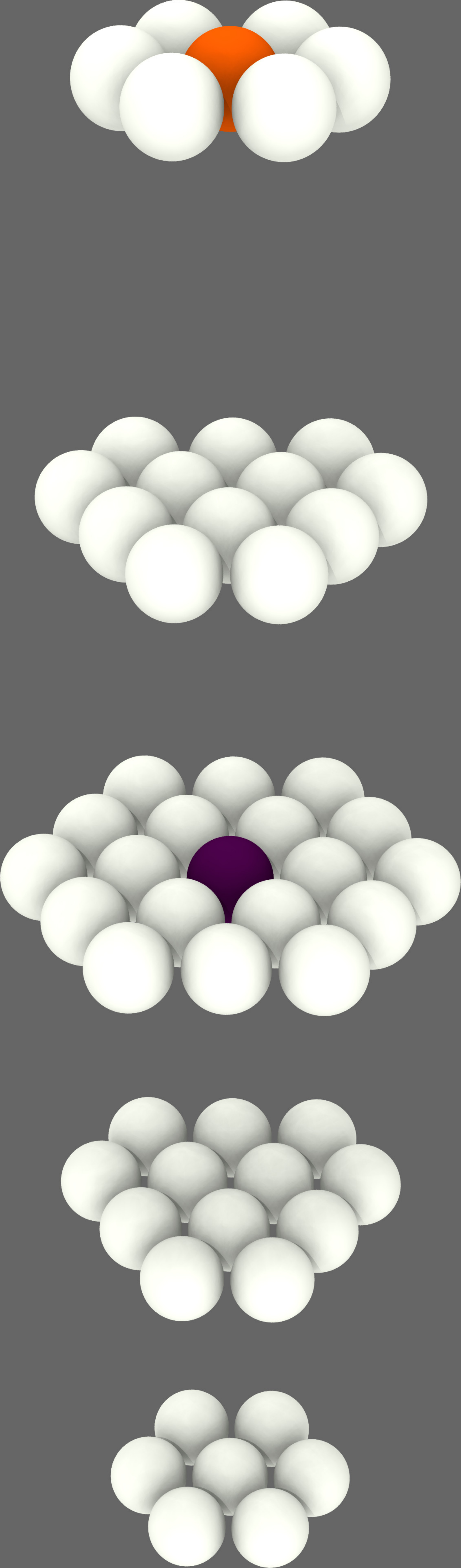}  
		\label{fig7r}
	}
	\caption{All Al environments around a vacancy found in SLKMC simulations (For color scheme and hop directions see Fig.~\ref{fig6}).}
	\label{patterns}
\end{figure*}

Interesting observation from Figs.~\ref{fig6}(a)-(c) and Table.~\ref{proc} is that the activation barriers for the in-plane exchange of vacancy and Mg atoms which are adjacent to the Al atom (P$_0$ and P$_4$ in Fig~\ref{fig6a}) are not equal to each other.  In Fig.~\ref{figa}(a), the exchange of Mg$_2$ \& V and Mg$_1$ \& V represent P$_0$ (0.283 eV) and P$_4$ (0.543 eV) in Fig~\ref{fig6a}, respectively.  Similarly, in Fig.~\ref{fig6c} activation barriers for seemingly symmetric vacancy hops P$_0$, P$_2$, P$_3$ and P$_5$ are not equal to each other.  But, in Fig.~\ref{fig6b} the activation barrier for P$_0$ is equal to P$_3$, and P$_5$ is equal to P$_6$ due to symmetry.  Seemingly symmetric vacancy hops in Fig.~\ref{fig6}(a) and (c) are in fact not symmetric due to the unequal distance between the saddle points of Mg atoms and the Al atom(s). 
%In Fig.~\ref{figa} one can see that the vacancy and Al atom in the diamond shape formed by Mg$_1$-Mg$_3$-V-Al are at the adjacent corners, while for the Mg$_1$-V-Mg$_3$-Al are along the short diagonal.
As noted earlier, within each ring a pair of two decimal ring numbers based on Al and Mg atom occupancy can represent the atomic neighborhood around Mg atoms. Accordingly, in Fig.~\ref{figa}(a) the neighborhoods of  Mg$_1$ and Mg$_2$  in the first ring are given as (1, 60) and (16, 39), respectively.  By performing symmetry operation, one can see that the atomic neighborhoods at initial location around Mg$_1$ and Mg$_2$ are not symmetric to each other. 
%Although, vacancy hops P$_0$ and P$_4$ appear symmetric, atomic neighborhoods seen by Mg$_1$ and Mg$_2$ during the hops are not symmetric due to the presence of the Al atom.  
%blue and red colored arrows show the paths taken by Mg$_1$ and Mg$_2$, respectively.   
In Fig.~\ref{figa} (a) one can see that at the saddle point Mg$_2$ is closer to Al atom than Mg$_1$.  Therefore, the activation barrier for P$_0$ is strongly influenced by the presence of the Al atom, whereas the activation barrier for P$_4$ is very close to the activation barrier for the vacancy-Mg atom exchange in pure Mg. Also, note that the size of the Al atom is smaller than the Mg atom.  For Mg$_1$ and Mg$_2$, atomic neighborhoods at the initial state and at the saddle point are different. Similar reasoning also applies for the asymmetry in the activation barriers for the vacancy hops; P$_0$, P$_2$, P$_3$ and P$_5$ shown in Fig.~\ref{fig6c}. 
%When the symmetry of the hop is viewed, it is important note that the Mg atoms on the adjacent layers should be also considered. 
%Moreover,  and hollow at the center of Mg$_2$-V-Al. Note that there also another atom in the layer above at the center of Mg$_1$-V-Al. 
%In Fig.~\ref{figa} one can also see that the V-Al pair in the diamond shape formed by Mg$_1$-Mg$_3$-V-Al and Mg$_1$-V-Mg$_3$-Al is on the adjacent corners and along the short diagonal, respectively. 
%In addition, in Fig.~\ref{figa} one can see an Mg atom (Mg$_0$) at the center of the triangle Mg$_1$-V-Al in the layer below, similarly there is also an Mg atom in the layer above. In case of the triangle Mg$_2$-V-Al, there is a hollow in both planes.

A single Al atom and a vacancy will produce a limited number of configurations.  A complete list of these Al atom configurations around a vacancy identified in SLKMC simulations of Al-atom diffusion in Mg matrix is shown in Fig.~\ref{patterns}, while corresponding vacancy-Mg and vacancy-Al exchange processes are given in Table.~\ref{All}. Note that for convenience Fig.~\ref{patterns} and Table.~\ref{All} show patterns and the activation barriers for vacancy-atom exchanges, respectively, for both A- and B-layers. One can see the asymmetry mentioned in the previous paragraph also in the activation barriers for vacancy hops within the basal plane as described in the previous paragraph.

\section{Conclusions}\label{conclusions}
On-lattice, self-learning KMC (SLKMC) simulations of Al atom diffusion in HCP Mg matrix are presented. SLKMC simulations take into account, both the effect of asymmetry of the HCP lattice  and the effect of asymmetry in the local atomic environment due to the presence of Al atom, on the activation barriers for vacancy-Mg and vacancy-Al atom exchange processes.  We presented a simple mapping method to map an HCP lattice on to a simple cubic lattice, which enables the study of non-cubic systems using on-lattice  framework. A comparison of  vacancy diffusivities in pure Mg matrix obtained from KMC simulations with those obtained from analytical expressions shows that the mapping accurately represents the HCP lattice. 

In agreement with the behavior that was observed both in theoretical and experimental studies, we also find the Al atom in Mg matrix diffuses faster with the basal plane than along the $c-$axis. The effective energy barriers differ from experimental and theoretical values by few tenths of an eV, but there seems to be a clear disagreement in the literature for the values of effective prefactors. Note that even though SLKMC simulations can identify new processes and calculate their activation barriers on the fly, not their pre-exponential factors which are, therefore, provided as an input. As pre-exponential factors do not vary significantly between different vacancy-atom exchange processes,  in our simulations they were assumed to be the same for all processes. 
Considering the fact that the MEAM potential shows good correlation with DFT and experimental values of materials parameters,  Al atom diffusivities obtained from SLKMC simulations are reasonable values. Importantly, more accurate effective activation energies and prefactors for Al atom diffusion can be obtained by performing KMC simulations using DFT calculated activation energies and pre-exponentials for vacancy-Mg and vacancy-Al exchange processes identified in present SLKMC simulations.

\section*{Acknowledgments}

This work was sponsored by the Vehicle Technologies office of U.S. Department of Energy (US DOE), Office of Energy Efficiency and Renewable Energy. A portion of this research was performed using computational resources at EMSL, a US DOE Office of Science, User Facility sponsored by the Office of Biological and Environmental Research, located at the Pacific Northwest National Laboratory (PNNL). PNNL is operated by Battelle Memorial Institute for the US DOE under DOE contract number DE-AC05-76RL1830. The authors would like to acknowledge the use of OVITO \cite{ovito} for visualization.

\section*{References}
\bibliography{references}

\providecommand{\newblock}{}
\begin{thebibliography}{10}
\expandafter\ifx\csname url\endcsname\relax
  \def\url#1{{\tt #1}}\fi
\expandafter\ifx\csname urlprefix\endcsname\relax\def\urlprefix{URL }\fi
\providecommand{\eprint}[2][]{\url{#2}}

\bibitem{Mg_ref_1}
Mordike B and Ebert T 2001 {\em Mater. Sci. Eng. A\/} {\bf 302}

\bibitem{Mg_ref_2}
Luo A~A 2003 {\em JOM\/} {\bf 54(2)}

\bibitem{Mg_ref_3}
Kulekci M 2008 {\em Int. J. Adv. Manuf. Technol\/} {\bf 39}

\bibitem{kristen}
Fichthorn K~A and Weinberg W~H 1991 {\em J. Chem. Phys.\/} {\bf 95} 1090

\bibitem{kmc1}
Amar J~G 2006 {\em Comput. Sci. Eng.\/} {\bf 8} 9

\bibitem{kmc_seg_1}
Clouet E and Soisson F 2010 {\em C. R. Phys.\/} {\bf 11} 226

\bibitem{kmc_seg_2}
Soisson F, Barbu A and Martin G 1996 {\em Acta. Mater.\/} {\bf 44} 3789

\bibitem{kmc_seg_3}
Bouar Y~L and Soisson F 2002 {\em Phys. Rev. B.\/} {\bf 65} 094103

\bibitem{kmc_seg_4}
Molnar D, Niedermeier C, Mora A, Binkele P and Schmauder S 2012 {\em Continuum
  Mech. Thermodyn.\/} {\bf 24} 607

\bibitem{kmc_seg_5}
Schmauder S and Binkele P 2002 {\em Comp. Mater. Sci.\/} {\bf 24} 42

\bibitem{kmc_seg_6}
Soisson F, Becquart C~S, Castin N, Domain C, Malerba L and Vincent E 2010 {\em
  J. Nucl. Mater.\/} {\bf 406} 55

\bibitem{kmc_seg_7}
Rautiainen T~T and Sutton A~P 1999 {\em Phys. Rev. B.\/} {\bf 59} 13681

\bibitem{kmc_seg_8}
Soisson F 2006 {\em J. Nucl. Mater.\/} {\bf 349} 235

\bibitem{kmc_seg_9}
Pareige C, Roussel M, Novy S, Kuksenko V, Olsson P, Domain C and Pareige P 2011
  {\em Acta. Mater.\/} {\bf 59} 2404

\bibitem{kmc_seg_10}
Vincent E, Becquart C~S, Pareige C, Pareige P and Domain C 2008 {\em J. Nucl.
  Mater.\/} {\bf 373} 387

\bibitem{kmc_seg_11}
Levesque M, Martinez E, Fu C~C, Nastar M and Soisson F 2011 {\em Phys. Rev.
  B.\/} {\bf 84} 184205

\bibitem{kmc_seg_12}
Martinez E 2012 {\em Phys. Rev. B.\/} {\bf 86} 224109

\bibitem{slkmc1}
Trushin O, Karim A, Kara A and Rahman T~S 2005 {\em Phys. Rev. B.\/} {\bf 72}
  115401

\bibitem{slkmc2}
Henkelman G and Jonsson H 2001 {\em J. Chem. Phys.\/} {\bf 115} 9657

\bibitem{a_slkmc2}
Nandipati G, Shim Y, Amar J~G, Karim A, Kara A, Rahman T~S and Trushin O 2009
  {\em J. Phys: Condens. Mater\/} {\bf 21} 084214

\bibitem{a_slkmc5}
Shah S~I, Nandipati G, Kara A and Rahman T~S 2012 {\em J. Phys: Condens.
  Mater\/} {\bf 24} 354004

\bibitem{a_slkmc7}
Nandipati G, Shah S~I, Kara A and Rahman T~S 2012 {\em J. Comput. Phys.\/} {\bf
  231} 3548

\bibitem{a_slkmc8}
Kara A, Trushin O, Yildirim H and Rahman T 2009 {\em J. Phys: Condens. Mater\/}
  {\bf 21} 084213

\bibitem{bkl}
Bortz A~B, Kalos M~H and Lebowitz J~L 1975 {\em J. Comput. Phys.\/} {\bf 17} 10

\bibitem{radix3}
Nandipati G and Kara A {\em (private communication)\/}

\bibitem{CINEB}
Henkelman G, Uberuaga B~P and Jonsson H 2000 {\em J. Chem. Phys.\/} {\bf 113}
  9901

\bibitem{LAMMPS}
Plimpton S 1995 {\em J. Comput. Phys.\/} {\bf 117} 1

\bibitem{meam1}
Lee B~J and Baskes M~I 2000 {\em Phys. Rev. B.\/} {\bf 62} 8564

\bibitem{meam2}
Lee B~J, Baskes M~I, Kim H and Cho Y~K 2001 {\em Phys. Rev. B.\/} {\bf 64}
  184102

\bibitem{mg_meam1}
Kim Y~M, Kim N~J and Lee B~J 2009 {\em CALPHAD\/} {\bf 33} 650

\bibitem{Mg_diff_dft}
Ganeshan S, Jr L~G~H and Liu Z~K 2010 {\em Comp. Mater. Sci.\/} {\bf 50} 301

\bibitem{Al_diff_dft}
Ganeshan S, Jr L~G~H and Liu Z~K 2011 {\em Acta. Mater.\/} {\bf 59} 3214

\bibitem{al_dft_new}
Zhou B~C, Shang S~L, Wang Y and Liu Z~K 2016 {\em Acta Metall.\/} {\bf 103} 573

\bibitem{all_diff_exp1}
Das S~K, Kim Y~M, Ha T~K, Gauvin R and Jung I~H 2013 {\em Metall. and Mater.
  Trans. A\/} {\bf 44} 2593

\bibitem{all_diff_exp11}
Das S~K, Kim Y~M, Ha T~K, Gauvin R and Jung I~H 2013 {\em Metall. and Mater.
  Trans. A\/} {\bf 44} 3420

\bibitem{all_diff_exp2}
Kammerer C~C, Kulkarni N~S, Warmack R~J and Sohn Y~H 2014 {\em J. Alloys and
  Compound\/} {\bf 617} 968

\bibitem{brennan}
Brennan S, Warren A~P, Coffey K~R, Kulkarni N, Todd P, Kilmove M and Sohn Y~H
  2012 {\em J. Phase Equilib. Diffus.\/} {\bf 33} 121

\bibitem{Brennen2}
Brennan S, Bermudez K, Kulkarni N~S and Sohn Y~H 2012 {\em Metall. and Mater.
  Trans. A\/} {\bf 43A} 4043

\bibitem{sd0}
Peterson N~L 1978 {\em J. Nucl. Mater.\/} {\bf 69-70} 3

\bibitem{sd1}
Mehrer H 2007 {\em Diffusion in Solids\/} (Berlin, Heidelberg: Springer)

\bibitem{sd2}
Combronde J and Brebec G 1971 {\em Acta Metall\/} {\bf 19} 1393

\bibitem{dft_barriers}
Mantina M 2008 {\em First-principles methodology for diffusion coefficients in
  metals and dilute alloys\/} Ph.D. thesis Pennsylvania State Univ.,

\bibitem{msd}
Einstein A 1905 {\em Ann. d. Phys\/} {\bf 17} 549

\bibitem{wagner}
Wagner C 1969 {\em Acta Metall.\/} {\bf 17 (2)} 99

\bibitem{BM1}
Boltzmann L 1894 {\em Ann. Phys. Chem. Wied.\/} {\bf 53} 959

\bibitem{BM2}
Matano C 1935 {\em Jpn. J. Phys.\/} {\bf 8} 109

\bibitem{Hall}
Hall L~D 1953 {\em J. Chem. Phys.\/} {\bf 21}

\bibitem{8freq}
Ghate P~B 1964 {\em Phys. Rev.\/} {\bf 133} A1167

\bibitem{13_freq}
Allnatt A~R, Belova I~V and Murch G~E 2014 {\em Phil. Mag.\/} {\bf 22} 2487

\bibitem{ovito}
Stukowski A 2009 {\em Modelling. Simul. Mater. Sci. Eng.\/} {\bf 18} 015012

\end{thebibliography}
 \end{document}